%% file: StringsBackground.tex
\documentclass[12pt]{utarticle}
\usepackage{amsmath,amsfonts,amssymb,bbm,mathrsfs}
\usepackage{epsfig,color}

\begin{document}

\title{\vskip-1cm Strings in Background Fields}
\author{Harun Omer}
\oneaddress{{\tt FirstnameLastname@gmail.com}}
\Abstract{In this paper free quantum theories are derived solely from their underlying symmetry group without reference to a Lagrangean
or classical physics and then interactions are introduced by making use of automorphism of the symmetry algebra. It is shown how the solution of a theory interacting with background fields is obtained exactly and purely algebraically from the free theory by the action of a linear operator. The method is first applied to the harmonic oscillator interacting with a conserved current and in a second example to string theory in the presence of constant dilaton, Kalb-Ramond or gravitational fields. The derived interaction for the Kalb-Ramond background is periodic and the interactions with a dilaton respectively a gravitational field exhibit a strong-weak-coupling duality. Furthermore, it is proposed in this paper to interpret T-duality as a position-momentum duality on compact space.}
\maketitle
\input{StringBgr_intro.tex}
\input{StringBgr_operator_algebra.tex}
\input{StringBgr_interactions.tex}
\input{StringBgr_exampleOsci.tex}
\input{StringBgr_examplePoincare.tex}
\input{StringBgr_exampleString.tex}
\input{StringBgr_appendix.tex}
\bibliographystyle{unsrt}   

\end{document}

%% file: StringBgr_intro.tex
\section{Introduction}
There is a need to revisit quantum theory and reformulate it in a way that avoids unnecessary assumptions and limitations.
Prime examples of quantum theories suspected to exist but impossible to formulate in practice are M-theory, F-theory and 6D superconformal theories. The latter are relevant for the description of the world-sheet of 7-branes in F-theory.
All three examples have no known action and it is not expected that Lagrangean formulations, in terms of which physicists usually work (and think), even exist. Moreover, as is advocated in a paper that is filed concurrently, string theory should ideally be formulated in a diffeomorphism invariant way, without the need of an early gauge-fixing. This paper avoids the use of Lagrangeans and the reference to classical physics. 
At every step it is attempted to keep assumptions to a minimum. For a free quantum theory a rather direct path from an underlying symmetry group to quantum theory is known, albeit hardly ever followed. Essentially one finds the unitary, irreducible projective representations of the space-time symmetry group.
The clearest exposition of this approach which I am aware of can be found in the reference~\cite{klink2015relativity}, which however restricts itself to the Galilean symmetry.
For relativistic quantum theory the  textbook closest to such an approach is~\cite{Weinberg:1995mt,Weinberg:1996kr}.
The identification of this underlying symmetry, under which the form of the laws of physics remains unchanged, is the first step towards formulating a quantum theory. A few relevant symmetry groups are collected in this table:\\\\
\begin{tabular}{c|c}
Symmetry Group & Significance\\
\hline
Galilei  & classical mechanics and so-called non-relativistic QM\\
Poincar\'e  & Einstein relativity and relativistic quantum theory\\
Super-Poincar\'e  & Poincar\'e invariance together with super-symmetry\\
2D (super-)conformal  & dynamics of the (supersymmetric) string world-sheet\\
6D superconformal & dynamics of a brane in F-theory
\end{tabular}
\\\\\\
This is only a tiny selection of symmetries which give rise to quantum theories of interest. In addition, slight modifications result in further physical theories. For instance introducing boundaries in 2D superconformal symmetry describes the physics of D-branes. The algebra underlying the Nambu-Goto or Polyakov action of the string is a coupling of the 2D conformal algebra and a higher-dimensional Poincar\'e algebra.

To derive the free quantum theory, we are looking for the unitary
irreducible projective representations of this Lie group.
Let us briefly sketch out this derivation, the details of which can be filled in from standard textbooks of quantum physics. If we postulate that quantum states obey the superposition principle, it implies that the states are elements of a vector space. The probability interpretation of quantum physics requires this space to be a topologically complete vector space with an inner product, that is, a Hilbert space. The presence of a symmetry means that the laws of physics remain form-invariant under a symmetry transformation. If an unprimed and a primed observer which are related by a symmetry transformation $g$ describe some quantum state by the vectors $|\phi\rangle$ and $|\phi'\rangle$ respectively, then an operator $U(g)$ must exist which relates the two states. By Wigner's theorem this operator must be linear (or anti-linear) and unitary (or anti-unitary).
In quantum theory we deal with projective representations since quantum states are rays rather than vectors in Hilbert space and consequently states are defined only up to a complex phase.
A projective unitary representation of a group $G$ is a mapping $U:G\mapsto GL(V)$ that satisfies the composition law of a linear mapping up to a phase.
 Since projective representations
are unwieldy to work with, in practice one works with the regular representations of the centrally extended symmetry group.
By a theorem of Bargmann~\cite{Bargmann:1954gh} the projective representation and the centrally extended regular representation are equivalent.
The central extension changes the algebra, for instance the commutator of
the space and momentum operators no longer commute~\cite{klink2015relativity},
\begin{eqnarray}
\;[\hat{X}_i,\hat{P}_j]=0\longrightarrow [\hat{X}_i,\hat{P}_j]=i\delta_{ij}.\nonumber
\end{eqnarray}
This is the quantization condition, which here is not imposed arbitrarily but results directly from the property that quantum states are defined as rays in Hilbert space.
Following our approach, one already has a one--"particle" Hilbert space together with observables and their commutation relations. Multi-"particle"\ states can be obtained by constructing the Fock space as a product space. The word "particle" has been placed in quotes since we may be dealing with objects with no particle interpretation in the case of string-theory. This procedure gives us a unique free theory from a given symmetry group. It has been attempted to stick so close to the basics of quantum theory that by loosening any assumptions, one quickly collides with the foundations of quantum theory, such as the superposition principle or the probability interpretation. The free theory can then be solved by algebraic means, such as analyzing the states of the physical space by finding a complete set of commuting observables (CSCO) and deriving states from the highest-weight states.
It must be noted that such a symmetry is not always sufficient to fully describe a quantum theory. The theory can be extended by introducing an additional internal symmetry groups such as for isospin, color and flavor symmetries.
The general approach is compatible with adding on such internal symmetries although conceptually it seems somewhat artificial. This issue can be partly overcome in supersymmetric theories. Supersymmetry is able to interweave an internal symmetry with a space-time symmetry.
String theory fully resolves this by geometrizing internal symmetries in one way or another. For instance in F-theory GUTs the gauge group of the standard model arises out of a geometric singularity.
Now let us turn to interacting quantum systems.
In contrast to the free theory, which is unique, many different interacting Lagrangeans can all lead to the same invariance algebra.
Conversely, the symmetry algebra does not suffice to define the full interacting theory. We can however ask for all possible interaction terms -- expressed directly in the operator formalism and without reference to an action -- which are consistent with the invariance algebra of the free theory. How this is done in practice is shown further below in this paper.
\\\\
\textbf{Acknowledgement:} I am very grateful to William Klink for helpful discussions and for him freely sharing his results.

%% file: StringBgr_operator_algebra.tex
\section{Operator Algebras}
\subsection{Invariance Algebra}
\subsubsection{Discrete Bosonic Operator Algebra}
For every Lie algebra exists an isomorphism between a matrix algebra representation and an operator algebra in terms of creation and annihilation operators. 
For simplicity, let us begin with a finite dimensional algebra. The idea is to express the $n \times n$ matrices of the representation in terms of the most elementary operators with non-trivial commutation relations, namely the bosonic creation and annihilation operators which satisfy the Weyl-Heisenberg algebra,
\begin{eqnarray}
\,[a_i,a^{\dagger}_j]=\delta_{ij}I \qquad [a_i,a_j]=0\qquad [a_i^{\dagger},a^{\dagger}_j]=0.
\end{eqnarray}
where $I$ is the identity operator and $i=1,...,n$. To construct a bosonic operator algebra, each matrix $A$ of the representation of the Lie algebra is assigned an equivalent operator $\mathcal{A}$:
\begin{eqnarray}
A \rightarrow \mathcal{A}_{bos}=a^{\dagger}Aa=\sum_{i,j}a_i^{\dagger}A_{ij}a_j.
\label{eq:bosalgeb}
\end{eqnarray}
The operator algebra inherits the commutation relations of the matrix algebra,
\begin{eqnarray}
\left[\mathcal{A}_{bos},\mathcal{B}_{bos}\right]=\mathcal{C}_{bos} \Leftrightarrow \left[A,B\right]=C,
\end{eqnarray}
which can be verified by evaluating the expression,
\begin{eqnarray}
\begin{array}{rcl}
\left[\mathcal{A}_{bos},\mathcal{B}_{bos}\right]&=&
\left[a_i^{\dagger} A_{ij} a_j , a_k^{\dagger} B_{kl} a_l\right]\\
&=&A_{ij} B_{kl} \left[a_i^{\dagger} a_j, a_k^{\dagger} a_l\right]\\
&=& A_{ij}B_{kl}\left(a_i^{\dagger}\delta_{jk}a_l-a_k^{\dagger}\delta_{li}a_j\right)\\
&=& a_i^{\dagger} \left[A,B\right]_{ij}a_j\\
&=&\mathcal{C}_{bos}.
\end{array}\label{eq:opalg}
\end{eqnarray}
Note that the isomorphism of the commutation relations also implies,
\begin{eqnarray}
e^{\mathcal{A}_{bos}}e^{\mathcal{B}_{bos}}=e^{\mathcal{C}_{bos}} \Leftrightarrow e^A e^B=e^C.
\end{eqnarray}
\subsubsection{Discrete Fermionic Algebra}
Instead of expanding into bosonic bilinears one can also expand into fermionic bilinears. In that case, the exact same isomorphism as for the bosonic algebra exists.  The canonical anti-commutation relationships for fermions is given by,
\begin{eqnarray}
\{b_i,b^{\dagger}_j\}=\delta_{ij}I\qquad \{b_i,b_j\}=0\qquad \{b_i^{\dagger},b^{\dagger}_j\}=0.
\end{eqnarray}
While the individual fermionic creation and annihilation operators satisfy anti-commutation relations, their bilinears satisfy commutation relations which are identical to the ones of the bosonic algebra,
\begin{eqnarray}
\begin{array}{ccc}
\,[b_i^{\dagger} b_j, b_k^{\dagger} b_l]&=&b_i^{\dagger}\delta_{jk}b_l-b_k^{\dagger}\delta_{li}b_j.
\end{array}
\end{eqnarray}
Since the isomorphism of the matrix algebra with the operator algebra relies only on this relationship, the proof of Eq.~(\ref{eq:opalg}) goes through unchanged for fermions. We can therefore construct a fermionic operator algebra in the same manner as a bosonic operator algebra:
\begin{eqnarray}
\left[\mathcal{A}_{fer},\mathcal{B}_{fer}\right]=\mathcal{C}_{fer} \Leftrightarrow \left[A,B\right]=C,
\end{eqnarray}
\begin{eqnarray}
e^{\mathcal{A}_{fer}}e^{\mathcal{B}_{fer}}=e^{\mathcal{C}_{fer}} \Leftrightarrow e^A e^B=e^C.
\end{eqnarray}
\subsubsection{Continuous Bosonic and Fermionic Operator Algebras}
The above argument is not restricted to a finite number of
basis states. It also readily generalizes to an infinite dimensional Lie algebra where the basis is continuous,
\begin{eqnarray}
a_i,a_i^\dagger,\sum_{i} \longrightarrow a(u),a^{\dagger}(u),\int du\nonumber,
\end{eqnarray}
The basis may also have both continuous and discrete labels. In the latter case the commutator is,
\begin{eqnarray}
[a(u,i),a^{\dagger}(u',j)]=i\delta_{ij}\delta(u-u')I,
\end{eqnarray}
and the mapping becomes,
\begin{eqnarray}
A \rightarrow \mathcal{A}=\sum_{i,j}\int \int a^{\dagger}(u,i)A(u,u',i,j)a(u',j)dudu'.
\end{eqnarray}
Again the same equivalence holds for fermions.
The continuous canonical commutation relationships can also be expressed in terms of a general scalar product,
\begin{eqnarray}
\;[a(u),a^{\dagger}(u')]=\langle u|u'\rangle
\qquad [a(u),a(u')]=0
\qquad [a^{\dagger}(u),a^{\dagger}(u')]=0.
\end{eqnarray}
All the results on discrete operator algebras in this work
can be readily applied to continuous operator algebras. Introducing interactions becomes more involved when the symmetry algebra acts on the continuous
parameter as is the case in relativistic quantum theory where the 
parameter is the four-momentum vector and the
Lorentz transformation acts on it. This paper does not
deal with such a case.
\subsubsection{Operator Algebras on $L^2$-space}
Isomorphisms to operator algebras on different Hilbert spaces also exist. 
The most prominent is the Hilbert space of square integrable functions $L^2$ with the elementary commutator bracket,
\begin{eqnarray}
\left[\partial_i,x_j\right]=\delta_{ij}.\label{eq:l2comm}
\end{eqnarray}
The operator algebra can then be defined by,
\begin{eqnarray}
\mathcal{A}=\sum_{i,j}  x_iA_{ij} \partial_j.
\end{eqnarray}
It is a standard procedure to start with a symmetry group and derive its generators in terms of coordinates $x_j$ and differentials $\partial_i$. Using the algebra in Eq.~(\ref{eq:l2comm}) we can express the generators in terms of creation and annihilation operators. But there is a sublety to take into account. Na\"ively one would be led to identify $\partial_i \simeq a_i$\ and $x_j \simeq a^{\dagger}_j$ as indeed this is a valid isomorphism.
However, $a_i$ and $a_i^{\dagger}$ are adjoint of each other whereas $\partial_i$ and $x_i$ are not.
The operator $x_i$ is self-adjoint since it is real.
The operator $\partial_i$ is skew-adjoint, which can be seen by integration by parts on the $L^2$-space, $\int \overline{f(x)}\frac{d}{dx}g(x)dx=-\int \frac{d\overline{f(x)}}{dx}g(x)dx.$ 
This deserves some more attention. After all, to derive  the physical content of the theory we will need unitary representations. We can obtain them from self-adjoint generators $A^{\dagger}=A$ or from skew-adjoint generators $B^{\dagger}=-B$ by exponentiation $U_{A}=e^{iAt}$ or $U_{B}=e^{ Bt}$. We could ask the question whether we can construct an algebra isomorphic to the Heisenberg-Weyl algebra out of two self-ajoint operators respectively two skew-adjoint operators. Let $\hat{x}_i^{\dagger}=\hat{x}_i$
and $\hat{p}_i^{\dagger}=\hat{p}_i$ denote self-adjoint operators. The analog of the commutator $[a_i,a_j^{\dagger}]$ would be $[\hat{x}_i,\hat{p}_j]$ which satisfies $[\hat{x}_i,\hat{p}_j]^{\dagger}=-[\hat{x}_i,\hat{p}_j]$. The same hold for skew-adjoint operators. To account for the sign-flip, the commutator must be imaginary:
\begin{eqnarray}
[\hat{x}_i,\hat{p}_j]=i\delta_{ij}I \qquad [\hat{x}_i,\hat{x}_j]=0\qquad [\hat{p}_i,\hat{p}_j]=0.
\end{eqnarray}
Due to the extra $i$ in the commutator an algebra defined by operators $\mathcal{A}=\sum_{i,j}\hat{x}_i A_{ij}\hat{p}_j$ would not satisfy Eq.~(\ref{eq:opalg}) and is not isomorphic to the matrix algebra. So the answer is no, we can not find an operator algebra in terms of two self-adjoint (or two skew-adjoint) operators. A way out is to work with one self-adjoint and one skew-adjoint operator,
\begin{eqnarray}
[\hat{x}_i,\frac{1}{i}\hat{p}_j]=\delta_{ij}I \qquad [\hat{x}_i,\hat{x}_j]=0\qquad [\hat{p}_i,\hat{p}_j]=0,
\end{eqnarray}
which brings us back to the algebra of Eq.~(\ref{eq:l2comm}).
Now let us get back to the original question to resolve the different behavior under Hermitian adjungation. We can identify,
\begin{eqnarray}
\begin{array}{rclrcl}
x_i&\simeq&\frac{1}{\sqrt{2}}(a_i^{\dagger}+a_i) & a_i&\simeq & \frac{1}{\sqrt{2}}(x_{i}-\partial_i)\\
\partial_i&\simeq&\frac{1}{\sqrt{2}}(a_i^{\dagger}-a_i) & a_i^{\dagger}&\simeq & \frac{1}{\sqrt{2}}(x_{i}+\partial_i).
\end{array}
\end{eqnarray}
Mathematically speaking, we are exploiting an automorphism of the algebra, applying a rotation in operator space until we have the desired properties under adjungation. 
Self-adjoint is the sum $a_i^{\dagger}+a_i$ and skew-adjoint the difference $a_i^{\dagger}-a_i$. This is equivalent to the more familiar identification known from the harmonic oscillator:
\begin{eqnarray}
\begin{array}{cc}
a_i=\frac{1}{\sqrt{2}}(\hat{x}_i+i\hat{p}_i)&\qquad a_i^{\dagger}=\frac{1}{\sqrt{2}}(\hat{x}_i-i\hat{p}_i)\\
\hat{x}_i=\frac{1}{\sqrt{2}}(a_i^{\dagger}+a_i) &\qquad \hat{p}_i=\frac{i}{\sqrt{2}}(a_i^{\dagger}-a_{i}).
\label{eq:oscipx0}
\end{array}
\end{eqnarray}
Automorphisms of the algebra will play an important role further below.
\subsubsection{Supersymmetric Operator Algebras}
In supersymmetric algebras one generally also has operators which are linear combinations of bilinears in one bosonic and one fermionic operator. Useful identities are,
\begin{eqnarray}
\begin{array}{rcl}
\left[\mathcal{A}_{bos},\mathcal{B}_{m}\right]&=&
\left[a_i^{\dagger} A_{ij} a_j , a_k^{\dagger} B_{kl} b_l\right]\\
&=&A_{ij} B_{kl} a_i^{\dagger}\left[a_j, a_k^{\dagger} \right]  b_{l}\\
&=& a_i^{\dagger} (A\cdot B)_{ij}b_j\\
&=&(\mathcal{AB})_{m},
\end{array}
\end{eqnarray}
and,
\begin{eqnarray}
\begin{array}{rcl}
\left\{\mathcal{A}_{m},\mathcal{B}_{fer}\right\}&=&
\left\{a_i^{\dagger} A_{ij} b_j , b_k^{\dagger} B_{kl} b_l\right\}\\
&=&A_{ij} B_{kl} a_i^{\dagger}\left\{b_j, b_k^{\dagger} \right\}b_{l}\\
&=& a_i^{\dagger} (A\cdot B)_{ij}b_j\\
&=&(\mathcal{AB})_{m}.
\end{array}
\end{eqnarray}
This will not be worked out in further detail here.
\subsection{Spectrum Generating Algebra}
I distinguish between the invariance algebra and the spectrum generating algebra\footnote{In the literature, spectrum generating algebra is sometimes used in a different sense.}. The invariance algebra reflects the invariance of the theory under the respective symmetry. It was argued above that the invariance algebra is generated by operators which are linear combinations of both a creation and an annihilation operator.
To generate the full spectrum we need the creation and annihilation operators $a_i^{\dagger}$ and $a_i$ separately. The algebra for the bosonic respectively fermionic creation and annihilation operators is given by,
\begin{eqnarray}
[a_i,a^{\dagger}_j]=\delta_{ij}I \qquad \{b_i,b^{\dagger}_j\}=\delta_{ij}I
\nonumber
\end{eqnarray}
or their continuous generalizations with all other brackets are vanishing. The identity operator $I$ is required for the algebra to close. If we enlarge the invariance algebra, which contains only bilinears, by adding the individual operators $a_i^{\dagger}$, $a_i$ (respectively $b_i^{\dagger}$, $b_i$) and the identity operator $I$, the enlarged algebra still closes. To introduce interactions, we will have to work with this enlarged algebra, which I call the spectrum-generating algebra.

%% file: StringBgr_interactions.tex
\section{Interactions}
\subsection{Properties of an Interacting Theory}
The irreducible unitary representations of the invariance algebra only define a free theory. Associated with a given symmetry are conserved quantities, such as the internal energy. The set of possible invariance transformations can map a given state only to a subset of the full Hilbert space, within the bounds of the conservation laws. Only within an interacting theory, the full eigenvalue spectrum of the CSCO is reachable. An interaction involves a sequence of actions of the operators $a_i^{\dagger},a_i,I$ of the spectrum generating algebra. In other words, we have to construct the interaction operators out of the operators of the spectrum generating algebra.
Therefore in an interacting theory, the generators $\mathcal{T}_f$ of the invariance algebra  acquire additional interaction terms $\mathcal{T}_I$ built out of operators of the spectrum-generating algebra,
\begin{eqnarray}
\mathcal{T}_f \rightarrow \mathcal{T}_f +g\mathcal{T}_I.
\end{eqnarray}
Here $g$ is a coupling constant which continuously deforms the free theory into an interacting theory. Consistency with the symmetry of the quantum system requires that the generators of the interacting theory continue to obey the invariance algebra of the free generators $T_f$. To preserve unitarity the interaction terms also have to be Hermitian.
The generators  of the free theory satisfy a Lie algebra,
\begin{eqnarray}
[\mathcal{T}^i,\mathcal{T}^j]=i f_{ijk}\mathcal{T}^k. \label{eq:liedef}
\end{eqnarray}
If we switch on an interaction term for some generator $\mathcal{T}^i$ and demand that the interacting operators $\mathcal{T}^i=\mathcal{T}^i_f+g \mathcal{T}^i_I$ still satisfy the invariance algebra,
then the mutual interrelation of all operators through the structure constants $f_{ijk}$ 
means that other operators will generally also have to acquire interaction terms. Depending on the structure of the interrelations,
these interaction terms can become rather complicated. In the following I am outlining how to obtain them systematically.
\subsection{Isomorphisms of the Algebra}
To simplify the problem, we first turn to the question of adding interaction terms to $a_i$ and $a_i^{\dagger}$ which preserve the Heisenberg-Weyl algebra and the unitarity condition. A deformation for which the operators remain Hermitian adjoint of each other is,
\begin{eqnarray}
\begin{array}{rcl}
a_j &\rightarrow& a_j+\lambda_j I\\
a_j^{\dagger} &\rightarrow& a_j^{\dagger}+\lambda_j^{*} I\\
\end{array}\qquad \lambda_j \in \mathbb{C}.\label{eq:subst}
\end{eqnarray}
This transformation is an automorphism of the bosonic algebra which means that the structure of its commutation relations is unaffected by it:
\begin{eqnarray}
\begin{array}{lcl}
\,[a_i+\lambda_iI,a_j^{\dagger}+\lambda_j^{*}I]&=&\delta_{ij}I,\\
\,[a_i+\lambda_iI,a_j+\lambda_jI]&=&0,\\
\,[a_i^{\dagger}+\lambda_i^{*}I,a_j^{\dagger}+\lambda_j^{*}I]&=&0.
\end{array} \nonumber
\end{eqnarray}
The generalization to continuous creation and annihilation operators is straightforward:
\begin{eqnarray}
\begin{array}{ccl}
[a(u,i) + \lambda(u,i)I,\,a^{\dagger}(u',j)+\lambda^*(u,j)I]&=&i\delta_{ij}\delta(u-u')I,\\\,
[a(u,i) + \lambda(u,i)I,\,a(u',j)+\lambda^*(u,j)I]&=&0,\\\,
[a^{\dagger}(u,i) + \lambda(u,i)I,\,a^{\dagger}(u',j)+\lambda^*(u,j)I]&=&0.
\end{array}
\nonumber
\end{eqnarray}
Since things work analogously in the continuous case it suffices to focus on the discrete case.  Given the transformed creation and annihilation operators, we can substitute them into any operator of the invariance algebra,
\begin{eqnarray}
\mathcal{A}(a^{\dagger}_{i},a_{i}) \rightarrow \mathcal{A}(a_i^{\dagger}+\lambda_i^{*}I,a_i+\lambda_i I).
\label{eq:defbosop}
\end{eqnarray}
Per Eq.~(\ref{eq:opalg}) all commutators of the invariance algebra are expressed in terms of the algebra of the creation and annihilation operators, so that all commutation relations remain unaffected by the substitution of Eq.~(\ref{eq:defbosop}). This is true generally, including for the commutation relations defining the invariance group Eq.~(\ref{eq:liedef}). Our transformation therefore also defines an automorphism of the invariance algebra. We have a set of operators parameterized by a set of complex numbers $\alpha_i$ (or a set of complex functions $\lambda_{i}(u)$) which act in the space of the extended algebra and satisfy the commutation relations of the invariance algebra.
Since automorphism preserve the structure of the defining algebra and as a result that of all operator expressions, it does not come as a surprise that the basic structure of the quantum theory is unaffected.
For instance, our free quantum theory has a CSCO given by
$\mathcal{A}^r:=\mathcal{A}^r(a_i^{\dagger},a_i)$, where the superscript $r$ labels the different operators of the set. Now suppose that for each operator exists a raising and a lowering operator,
\begin{eqnarray}
\begin{array}{ccl}
\,[\mathcal{A}^r(a_i^{\dagger},a_i),\mathcal{R}^r(a_i^{\dagger},a_i)]&=&+\lambda^r \mathcal{R}^r(a_i^{\dagger},a_i)\\
\,[\mathcal{A}^r(a_i^{\dagger},a_i),\mathcal{L}^r(a_i^{\dagger},a_i)]&=&-\lambda^r\mathcal{L}^r(a_i^{\dagger},a_i).
\end{array}\nonumber
\end{eqnarray}
Then the same relation is preserved under the transformation and the interacting theory is characterized by an  eigenvalue spectrum isomorphic to the free theory. The proposed substitution corresponds to the 'primitive interaction terms' of~\cite{klinkunpub}. To find more general interaction terms we need
to work with functions of the creation and annihilation operators, for which the following identities are useful:
\begin{eqnarray}
\begin{array}{rclrcl}
\;[a,a^{\dagger k}]&=&k a^{\dagger k-1} &
\;[a^{\dagger},a^k]&=&-ka^{k-1}\\
\,[a,a^{\dagger k}a^l]&=&k a^{\dagger k-1}a^l &
\; [a^{\dagger},a^k a^{\dagger l}]&=&-k a^{k-1}a^{\dagger l}.
\end{array}\label{eq:opderiv}
\end{eqnarray}
Further identities can be found in~\cite{jpain2012}. From the above relations we conclude,
\begin{eqnarray}
[a,f(a^{\dagger},a)]=\partial_1f(a^{\dagger},a) \qquad[a^{\dagger},f(a^{\dagger},a)]=\partial_2f(a^{\dagger},a).
\end{eqnarray}
When we deform by a general function of the creation and annihilation operators,
\begin{eqnarray}
[a+f(a^{\dagger},a),a^{\dagger}+f(a^{\dagger},a)^{\dagger}]\overset{!}{=}1 \nonumber
\end{eqnarray}
we obtain the condition,
\begin{eqnarray}
\partial_1f(a^{\dagger},a) -\partial_2f(a^{\dagger},a) + [f(a^{\dagger},a),f(a^{\dagger},a)^{\dagger}]=0,
\end{eqnarray}
which is satisfied for $f(a^{\dagger},a)=iv(a^{\dagger},a)$ with $v(a^{\dagger},a)^{\dagger}=v(a^{\dagger},a)$. More deformations are discussed further below. In principle one can proceed in the same way with the fermionic algebra. The fermionic operator algebra is given by,
\begin{eqnarray}
\{b_i,b_j^{\dagger}\}=\delta_{ij}I\qquad \{b_i,b_j\}=0 \qquad \{b_i^{\dagger},b_j^{\dagger}\}=0.
\end{eqnarray}
The anti-commutators make it more difficult to find deformations of the operators preserving the algebra. In particular, the fermionic equivalence of the simplest bosonic transformation would be,
\begin{eqnarray}
\begin{array}{rcl}
b_j &\rightarrow& b_j+\gamma_j I\\
b_j^{\dagger} &\rightarrow& b_j^{\dagger}+\gamma_j^{*} I,\\
\end{array}\nonumber
\end{eqnarray}
but this deformation does not satisfy the original algebra for any $\gamma_j \in \mathbb{C}$. One would have to elevate the $\gamma_j$ to Grassmann variables to have a valid deformation. 
\subsection{Solving the Interacting Theory}
\subsubsection{Isomorphic Algebras from Unitary Transformations}
We can find isomorphism of the algebra with the help of unitary operators.
This method does not appear to have been used in the literature widely for this purpose. However, Witten in his paper on Morse theory\cite{Witten:1982im} used it in a similar way by applying it on the exterior derivative of a supersymmetric algebra in order to obtain deformed theories. That reference uses the transformation,
\begin{eqnarray}
d^{\dagger}_t=e^{ht}d^{\dagger}e^{-ht}\qquad d_t=e^{-ht}de^{ht},
\end{eqnarray}
which preserves the supersymmetric algebra. Here $t$ is a parameter and $h$ a function. In this work I am generalizing the transformation to one which itself may depend on creation and annihilation operators.
Consider an operator $U$ which acts on the annihilation operators,
\begin{eqnarray}
a_i \rightarrow Ua_iU^{-1}\nonumber
\end{eqnarray}
Typically $U$ is the function of a continuous parameter so that $U$ deforms $a_i$ continuously. The adjoint action is,
\begin{eqnarray}
a^{\dagger}_i \rightarrow U^{-1 \dagger}a^{\dagger}_iU^{\dagger} \nonumber
\end{eqnarray}
We can multiply any relation of the algebra from the left with $U$ and from the right with $U^{-1}$ and insert the identity operator $U^{-1} U=I$ between operators, for instance,
\begin{eqnarray}
(Ua^{\dagger}_iU^{-1})\ (U a_jU^{-1})-(Ua_j U^{-1})(Ua^{\dagger}_iU^{-1}) =\delta_{ij}I.\nonumber
\end{eqnarray}
When $U$ is a unitary operator, $U^{-1}=U^{\dagger}$,
we found an isomorphic algebra where the creation and annihiliation operators continue to be Hermitian adjoint of each other.
The application to the fermionic algebra works in the same manner. 
\subsubsection{From the Free Theory to the Solution of the Interacting Theory}
Suppose we have transformed the free theory by virtue of a unitary
operator $U$ to an interacting theory, or any deformed theory for that matter.
This deformation then ascends to a deformation of the full symmetry algebra of the system. Any operator $\mathcal{A}_f$ of the free theory is related
to an operator $\mathcal{A}_I$ of the interacting theory by,
\begin{eqnarray}
\mathcal{A}_I = U\mathcal{A}_fU^{-1}.
\end{eqnarray}
Similarly, free and interacting states in the Hilbert space are interrelated by,
\begin{eqnarray}
|\Psi_I\rangle = U|\Psi_f\rangle.
\end{eqnarray}
The solution of the free theory and knowledge of the operator $U$ is therefore sufficient to solve the interacting theory. By construction, such interacting theories are always unitarily equivalent to some free theory. But while for a unitary transformation based on the symmetry algebra one has
$U_{\text{inv}}|0\rangle=|0\rangle$, this is not the case for the interacting operators.
They affect the vaccuum state and thereby describing background fields. Note that the basic idea to modify the free generators in a way that the invariance algebra continues to be satisfied is not limited to theories which are unitarily equivalent to a free theory.
\subsubsection{Connection to the Heisenberg Picture}
Define a function,
\begin{eqnarray}
\hat{F}(q)=e^{q\hat{X}} \hat{Y} e^{-{q\hat{X}}},
\end{eqnarray}
where $\hat{X}$ and $\hat{Y}$ are operators and $q$ is a c-number. Differentiating this equation gives a differential equation in the form of the equation of motion
in the Heisenberg picture:
\begin{eqnarray}
\frac{d}{dq}\hat{F}(q)=[\hat{X},\hat{F}(q)].
\end{eqnarray}
We can insert a series expansion of $\hat{F}(q)$ into the equation,
\begin{eqnarray}
\sum_{i=1}^{\infty}\frac{1}{(i-1)!}\hat{F}_iq^{i}=\sum_{i=0}^{\infty}\frac{1}{i!}[\hat{X},\hat{F}_i]q^i,
\end{eqnarray}
and solve it iteratively. Then $\hat{F}_0=\hat{Y}$ and $\hat{F}_{i+1}=[\hat{Y},\hat{F}_i]$.
Reconstituting the series expansion and setting $q=1$ we recover the Baker-Hausdorff Lemma,
\begin{eqnarray}
e^{\hat{X}}\hat{Y}e^{-\hat{X}}=\hat{Y}+[\hat{X},\hat{Y}]+\frac{1}{2!}[\hat{X},[\hat{X},\hat{Y}]]+\frac{1}{3!}[\hat{X},[\hat{X},[\hat{X},\hat{Y}]]]+ \dots\label{eq:bakerhausdorff}
\end{eqnarray}
\subsection{Deformation Operators}
This paragraph returns to discussing isomorphisms of the invariance algebra. The idea is to establish the action of a unitary
operator $U=e^{\hat{X}}$ on the creation and annihilation
operators as the elemetary building
blocks of the quantum theory from which we can derive their action on any operator. As mentioned before, this will enable us to solve the
deformed theory given the solution of the free theory. In the computation we will frequently resort to the Baker-Hausdorff Lemma.\\\\
\underline{$\hat{X}= i \phi$:}\\
Conjugation by a complex phase $e^{i\phi}$ acts trivially on the
bosonic and fermionic operators:
\begin{eqnarray}
\begin{array}{rclrcl}
e^{i\phi}a_je^{-i\phi} = a_j & \qquad e^{i\phi}b_je^{-i\phi} = b_j \\
e^{i\phi}a_j^{\dagger}e^{-i\phi} = a_j^{\dagger} & \qquad e^{i\phi}b_j^{\dagger}e^{-i\phi} = b_j^{\dagger}
\end{array}
\qquad \phi \in \mathbb{R}.
\end{eqnarray}
\\\\
\underline{$\hat{X}=\lambda a^{\dagger}-\lambda^* a$:}\\
The simplest non-trivial transformation which leaves the bosonic algebra invariant is the one discussed earlier:
\begin{eqnarray}
\begin{array}{rcl}
a_j &\rightarrow& a_j-\lambda_j I\\
a_j^{\dagger} &\rightarrow& a_j^{\dagger}-\lambda_j^{*} I\\
\end{array}\qquad \lambda_j \in \mathbb{C}.\nonumber
\end{eqnarray}
The goal is to  find a unitary operator which transforms the operators in the desired way by conjugation. That means we are looking for an operator with the property,
\begin{eqnarray}
\begin{array}{rcl}
D(\lambda)a_{j}D(\lambda)^{-1}&=&a_{j}-\lambda_j\\
D(\lambda)a^{\dagger}_{j}D(\alpha)^{-1}&=&a^{\dagger}_j-\lambda^*_j.
\end{array}
\end{eqnarray}
Unitarity requires $D(\lambda)^{\dagger}=D(\lambda)^{-1}=D(-\lambda)$.
If $D(\lambda)$ is a solution, the operator multiplied by a phase $e^{ic}$ will also solve the equation. By rearranging the terms we see that $D(\lambda)$ is a type of raising respectively lowering operator:
\begin{eqnarray}
\begin{array}{ccl}
\,[a_k,D(\lambda)]&=&D(\lambda)\lambda_k\\
\,[a^{\dagger}_k,D(\lambda)]&=&D(\lambda)\lambda^*_k.\nonumber
\end{array}
\end{eqnarray}
Such an operator is easy to find. A general anti-Hermitian element of the algebra can be written,
\begin{eqnarray}
icI +\sum_{i}(\lambda_{i}a^{\dagger}_i-\lambda^*_{i}a_{i}) \qquad c\in \mathbb{R}, \lambda \in \mathbb{C}^n,\nonumber
\end{eqnarray}
from which we obtain an irreducible unitary representation by exponentiation,
\begin{eqnarray}
U=e^{ic}D(\lambda).\nonumber
\end{eqnarray}
The operator $D(\lambda)$ is  given by,
\begin{eqnarray}
D(\lambda):=\exp\{\sum_i \lambda_i a_i^{\dagger}-\lambda_i^*a_i\}.
\end{eqnarray}
It has the desired action on the creation and annihiliation operators. Under multiplication we have,
\begin{eqnarray}
D(\lambda)D(\lambda')=D(\lambda+\lambda')e^{i\text{Im}(\lambda\lambda^*)}.
\end{eqnarray}
The action on the vacuum state gives,
\begin{eqnarray}
|\lambda\rangle \equiv D(\lambda)|0\rangle = e^{-\frac{1}{2}\sum_i\lambda_i^*\lambda_i}e^{\sum_i\lambda_i a_i^{\dagger}}|0\rangle.
\end{eqnarray}
The resulting states are eigenstates of the annihilation operators,
\begin{eqnarray}
a_i|\lambda\rangle =\lambda_i |\lambda\rangle.
\end{eqnarray}
The operator is familiar from the treatment of the harmonic oscillator and is called displacement operator. The displacement operator derives its name from displacing the ground state in the $a_i,a_i^{\dagger}$-space or -- equivalently for the harmonic oscillator -- in the $\hat{X},\hat{P}$-space.
The states $|\lambda\rangle$ are called coherent states 
which are familiar in particular from quantum optics~\cite{Glauber:1963fi,Glauber:1963tx}. \\\\
\underline{$\hat{X}= \beta b^{\dagger}+\beta^* b$:}\\
It has been mentioned before that the deformation of the bosonic operators by a constant has a fermionic equivalent only in terms of Grassmann numbers. Then the displacement operator for the fermions has the same structure as that for the bosons. Such an operator has appeared in the literature before~\cite{PhysRevA.59.1538}, but it does not seem to have been applied widely with success.  Here we restrict to ordinary c-numbers and compute the action of the operator from the Baker-Hausdorff Lemma in Eq.~(\ref{eq:bakerhausdorff}).
The relevant commutators in the expansion are,
\begin{eqnarray}
\begin{array}{rcl}
\,[\tilde{\beta} b +\beta b^{\dagger},\tilde{\beta} b-\beta b^{\dagger}]&=&2(\beta\tilde{\beta})(2b^{\dagger}b-1)\\
\,[\tilde{\beta} b +\beta b^{\dagger},2b^{\dagger}b-1]&=&2(\tilde{\beta} b-\beta b^{\dagger})
\end{array}
\end{eqnarray}
Using them one gets,
\begin{eqnarray}
\begin{array}{rcl}
e^{\tilde{\beta} b +\beta b^{\dagger}}be^{\tilde{\beta} b -\beta b^{\dagger}}&=&b+2\beta(\tilde{\beta} b-\beta b^{\dagger})\left\{\frac{1}{2!}+\frac{(4\beta\tilde{\beta})}{4!}+\frac{(4\beta\tilde{\beta})^2}{6!}+\dots\right\}
\\
&&-\beta (2b^{\dagger}b-1)\left\{\frac{1}{1!}+\frac{(4\beta\tilde{\beta})}{3!}+\frac{(4\beta\tilde{\beta})^2}{5!}+\dots \right\}\\
&=&b+\frac{1}{2\tilde{\beta}}(\tilde{\beta} b-\beta b^{\dagger})(\cosh(2\sqrt{\beta\tilde{\beta}})-1)-\frac{\sqrt{\beta}}{2\sqrt{\beta}}(2b^{\dagger}b-1)\sinh(2\sqrt{\beta\tilde{\beta}})
\end{array}\nonumber
\end{eqnarray} 
so that,
\begin{eqnarray}
\begin{array}{l}
e^{\tilde{\beta} b +\beta b^{\dagger}}be^{-\tilde{\beta} b -\beta b^{\dagger}}=b+\frac{1}{2\tilde{\beta}}(\tilde{\beta} b-\beta b^{\dagger})(\cosh(2\sqrt{\beta\tilde{\beta}})-1)-\frac{\sqrt{\beta}}{2\sqrt{\tilde{\beta}}}(2b^{\dagger}b-1)\sinh(2\sqrt{\beta\tilde{\beta}})\\
e^{\tilde{\beta} b +\beta b^{\dagger}}b^{\dagger}e^{\tilde{-\beta} b -\beta b^{\dagger}}=b^{\dagger}-\frac{1}{2\beta}(\tilde{\beta} b-\beta b^{\dagger})(\cosh(2\sqrt{\beta\tilde{\beta}})-1)+\frac{\sqrt{\tilde{\beta}}}{2\sqrt{\beta}}(2b^{\dagger}b-1)\sinh(2\sqrt{\beta\tilde{\beta}}).\\ \end{array}
\end{eqnarray} 
The operator is unitary when $\tilde{\beta}= \beta^*$.\\\\
\underline{$\hat{X}=\frac{1}{2}\xi a^{\dagger 2} - \frac{1}{2}\xi^* a^2$:}\\
We can define the operator,
\begin{eqnarray}
S(\xi)=\exp\{\frac{1}{2}\xi a^{\dagger 2} - \frac{1}{2}\xi^* a^2\},
\end{eqnarray}
which is unitary and satisfies $S(\xi)^{\dagger}=S(\xi)^{-1}=S(-\xi)$.
Its action on the creation and annihilation operators is,
\begin{eqnarray}
\begin{array}{ccl}
S(\xi)aS(\xi)^{-1}&=& a \cosh(s) -a^{\dagger}e^{i\theta}\sinh(s)\\
S(\xi)a^{\dagger}S(\xi)^{-1}&=&a^{\dagger}\cosh(s) -  a e^{-i\theta}\sinh(s)
\end{array}\text{ with } \xi =s e^{i\theta}.
\end{eqnarray}
The operator is also familiar from quantum optics 
where it is known as squeezing operator. Just like for coherent states, for
squeezed states the amplitude-phase uncertainty bound is saturated. For coherent states the uncertainty region is circular, for squeezed states the circle is "squeezed" to an ellipse, reducing the amplitude uncertainty at the expense of the phase uncertainty or vice versa, hence its name.
Due to the nilpotence of the fermionic ladder operators the equivalent
fermionic operator reduces to the identity. But states with squeezed uncertainty region exist also for fermions~\cite{PhysRevD.43.1403,Nieto:242326}.
\\\\
\underline{$\hat{X}= i\sum_{k}\sum_{l=0}^{\infty} \frac{1}{(l+1)}g^{l}_{j,k} \left(e^{-i\pi \phi_{j,k}} a^{\dagger}_k +e^{i\pi \phi_{j,k}} a_k\right)^{l+1}$:}\\
We can define a deformation operator,
\begin{eqnarray}
D(g,\phi)=\exp\left\{i\displaystyle\sum_{k}\sum_{l=0}^{\infty} \frac{1}{(l+1)}g^{l}_{j,k} \left(e^{-i\pi \phi_{j,k}} a^{\dagger}_k +e^{i\pi \phi_{j,k}} a_k\right)^{l+1}\right\}\;\; g_k^l \in \mathbb{R}\;\; \phi_{j,k}\in [0,2\pi],
\end{eqnarray}
which acts on the creation and annihilation operators according to,
\begin{eqnarray}
\begin{array}{rcl}
D(g,\phi) a_{j}D^{-1}(g,\phi)&=&a_j\,-i\displaystyle\sum_{k}\sum_{l=0}^{\infty} g^{l}_{j,k} e^{-i\pi \phi_{j,k}}\left(e^{-i\pi \phi_{j,k}} a^{\dagger}_k +e^{i\pi \phi_{j,k}} a_k\right)^l \\
D(g,\phi){a_j}^{\dagger}D^{-1}(g,\phi)&=&a_j^{\dagger}+i\displaystyle\sum_{k}\sum_{l=0}^{\infty}g_{j,k}^{l}e^{+i\pi \phi_{j,k}}\left(e^{-i\pi \phi_{j,k}} a^{\dagger}_k +e^{i\pi \phi_{j,k}} a_k\right)^l.
\end{array}
\end{eqnarray}
The operator is unitary and satisfies $D^{-1}(g,\phi)=D(g,\phi)^{\dagger}=D(-g,\phi)$.
Note that the deformation is essentially an arbitrary smooth function  which takes the
self-adjoint combinations $e^{-i\pi \phi_{j,k}} a^{\dagger}_k +e^{i\pi \phi_{j,k}} a_k$ as arguments. The exponent in $D(g,\phi)$ is the integral of the function. Therefore this is the case which has been discussed in the earlier section about isomorphisms of the algebra after Eq.~(\ref{eq:opderiv}).\\\\
\underline{$\hat{X}=M_{ij}a^{\dagger}_ia_j$:}\\
Here the exponent of the unitary operator is a sum of bilinears. If (and only if) they are all part of the invariance algebra the transformation reduces to a symmetry.
We have $[\hat{X},a_k]=-M_{kj}a_j$ and $[\hat{X},a^{\dagger}_k]=M_{ik}a^{\dagger}_i$, so we obtain:
\begin{eqnarray}
\begin{array}{ccl}
e^{a^{\dagger}_iM_{ij}a_j}a_ke^{-a^{\dagger}_iM_{ij}a_j}&=&(e^{-M})_{kj}a_{j},\\
e^{a^{\dagger}_iM_{ij}a_j}a^{\dagger}_ke^{-a^{\dagger}_iM_{ij}a_j}&=&(e^{M^{T}})_{kj}a^{\dagger}_{j}.
\end{array}\label{eq:bilineartrans}
\end{eqnarray}
As always, the exponentiated operator $e^{\hat{X}}$ is unitary when $\hat{X}$ is anti-Hermitian. $\hat{X}=M_{ij}a^{\dagger}_ia_j$ is anti-Hermitian when $M$ is. One can define $R$ through $M=\ln R$ so that $(e^{M})_{kj}a_{j}=R_{kj} a_j$. When $B$ is real so that anti-hermiticity reduces to anti-symmetry, $M=-M^T$, then $R$ and can be interpreted as a type of rotation matrix. For a diagonal operator  $M_{ij}=i\theta \delta_{ij}$ one has $\hat{N}\equiv\hat{X}=i\theta \sum_i a^{\dagger}_ia_i$ so that,
\begin{eqnarray}
\begin{array}{ccc}
e^{i\theta \hat{N}}a_ke^{-i\theta \hat{N}}&=&e^{-i\theta}a_{k},\\
e^{i\theta \hat{N}}a^{\dagger}_ke^{-i\theta \hat{N}}&=&e^{i\theta}a^{\dagger}_{k}.
\end{array}
\end{eqnarray}
where $\theta$ is a real phase to ensure unitarity of the deformation operator.\\\\
\underline{$\hat{X}=M_{ij}b^{\dagger}_ib_j$:}\\
The structure of the commutators $[\hat{X},b_k]=-M_{kj}b_j$ and $[\hat{X},b^{\dagger}_k]=M_{ik}b^{\dagger}_i$ is identical to the bosonic case and as a result the conjugation acts in the same way:
\begin{eqnarray}
\begin{array}{ccl}
e^{b^{\dagger}_iM_{ij}b_j}b_ke^{-b^{\dagger}_iM_{ij}b_j}&=&(e^{-M})_{kj}b_{j},\\
e^{b^{\dagger}_iM_{ij}b_j}b^{\dagger}_ke^{-b^{\dagger}_iM_{ij}b_j}&=&(e^{M^{T}})_{kj}b^{\dagger}_{j}.
\end{array}
\end{eqnarray}
The fermionic operator $\hat{X}=M_{ij}b^{\dagger}_ib_j$ is anti-Hermitian when $M$ is Hermitian. Again one can define $R$ through $M=-\ln R$ so that $(e^{-M})_{kj}a_{j}=R_{kj} a_j$ where $R$ is Hermitian. For a diagonal operator  $M_{ij}=\phi \delta_{ij}$ the transformation is a rescaling,
\begin{eqnarray}
\begin{array}{ccc}
e^{\phi \hat{N}}b_ke^{-\phi \hat{N}}&=&e^{-\phi}b_{k},\\
e^{\phi \hat{N}}b^{\dagger}_ke^{-\phi \hat{N}}&=&e^{\phi}b^{\dagger}_{k}.
\end{array}
\end{eqnarray}
The scaling factor $\phi$\ needs to be real in order for the operator to be unitary.\\\\
\underline{$\hat{X}=\frac{1}{2}M_{ij}(a^{\dagger}_ia^{\dagger}_j-a_ia_j)$:}\\
For non-vanishing $\hat{X}$ the matrix $M_{ij}$ must be symmetric. We further take $M_{ij}$ to be real and obtain the deformation,
\begin{eqnarray}
\begin{array}{ccr}
e^{\frac{1}{2}a^{\dagger}_iM_{ij}a^{\dagger}_j-\frac{1}{2}a_iM_{ij}a_j}a_ke^{-\frac{1}{2}a^{\dagger}_iM_{ij}a^{\dagger}_j+\frac{1}{2}a_iM_{ij}a_j}&=& \cosh(M)_{kl}a_{l}-\sinh(M)_{kl}a^{\dagger}_l,\\
e^{\frac{1}{2}a^{\dagger}_iM_{ij}a^{\dagger}_j-\frac{1}{2}a_iM_{ij}a_j}a^{\dagger}_ke^{-\frac{1}{2}a^{\dagger}_iM_{ij}a^{\dagger}_j+\frac{1}{2}a_iM_{ij}a_j}&=&-\sinh(M)_{kl}a_{l} +\cosh(M)_{kl}a^{\dagger}_l.
\end{array}
\end{eqnarray}
\\\\
\underline{$\hat{X}=\frac{1}{2}M_{ij}(b^{\dagger}_ib^{\dagger}_j-b_i b_j)$:}\\
The fermionic case is opposite to the bosonic case in the sense that the symmetric contributions drop out, so we take the matrix to be anti-symmetric $M_{ij}=-M_{ji}$. Using $[X,b_k]=-M_{kj}b^{\dagger}_j$ and $[X,b^{\dagger}_k]=M_{kj}b_{j}$ one finds,
\begin{eqnarray}
\begin{array}{ccl}
e^{\frac{1}{2}M_{ij}(b^{\dagger}_ib^{\dagger}_j-b_i b_j)}b_ke^{-\frac{1}{2}M_{ij}(b^{\dagger}_ib^{\dagger}_j-b_i b_j)}&=&\cos (M)_{kl}b_l-\sin(M)_{kl}b^{\dagger}_l,\\
e^{\frac{1}{2}M_{ij}(b^{\dagger}_ib^{\dagger}_j-b_i b_j)}b^{\dagger}_ke^{-\frac{1}{2}M_{ij}(b^{\dagger}_ib^{\dagger}_j-b_i b_j)}&=&\cos(M)_{kl}b^{\dagger}_l+\sin(M)_{kl}b_l.
\end{array}
\end{eqnarray}
It may be useful to define the operators using different ordering conventions.
A normal ordered version of the above unitary operator would be,
\begin{eqnarray}
:e^{\frac{1}{2}M_{ij}(b^{\dagger}_ib^{\dagger}_j-b_i b_j)}:
=e^{\frac{1}{2}M_{ij}b^{\dagger}_ib^{\dagger}_j}e^{-\frac{1}{2}M_{ij}b_ib_j},\nonumber
\end{eqnarray}
and similarly for the bosonic operators.
\subsubsection{Supersymmetric Deformation Operators}
It is possible to construct deformation operators $e^{\hat{X}}$ which contain both bosonic and fermionic creation and annihilation operators.
In principle they can be found in a similar manner although 
their action will generally not simplify as much as in the pure
bosonic or fermionic cases.
\subsubsection{S-Matrix}
Since the procedure proposed above deals with various types of coherent states, a few remarks on them are in order.
The physically measurable quantities in quantum field theories are derived from vacuum to vacuum transition amplitudes in the presence of an external source. When the interaction with the external source is linear, the final state is a coherent state. In the computation of scattering amplitudes one assumes the interaction to be switched on adiabatically, so that the incoming and outgoing states are free fields. The sets of incoming and outgoing states are complete sets of free states on the Fock space created by free field operators. Two such complete sets must be related to one another by a unitary transformation,
\begin{eqnarray}
\begin{array}{rcl}
A_{\text{out}} &=& S^{\dagger}A_{\text{in}}S,\\
|\text{out}\rangle &=& S^{\dagger}|\text{in}\rangle
\end{array}
\end{eqnarray}
This transformation matrix is called the S-matrix.
For more on this see for instance~\cite{Zhang:1999is}.

%% file: StringBgr_exampleOsci.tex
\section{Quantum Theory from the $U(n)$ and $SU(n)$ Groups}
\subsection{Classical Symmetry of the Harmonic Oscillator}
The group $U(n)$ -- as well as its subgroup $SU(n)$ -- preserves the standard inner product on $\mathbb{C}^n$:
\begin{eqnarray}
z^{\dagger}M^{\dagger}Mz=z^{\dagger}z=\text{const}\qquad M \in U(n).\nonumber
\end{eqnarray}
The complex vector $z$ can be expanded into its real and imaginary components,
\begin{eqnarray}
z_i=x_i+ip_i \qquad z^{\dagger}z=\vec{p}^2+\vec{x}^2=\text{const}.\nonumber
\end{eqnarray}
The $U(n)$ symmetry completely determines the phase space of the harmonic oscillator.
Whether we start out with unitary or special unitary invariance is irrelevant for the quantum theory since the projective unitary group $PU(n)$ is identical to the projective special unitary group $PSU(n)$ and in both cases a central extension leads to a vector representation of $U(n)$. By working merely with a $U(n)$ symmetry, the time and energy dependence is ignored. This could be easily remedied by working with a different group (or a tensor product of groups), but it is instructive to restrict to the simpler $U(n)$ as a first example.
\subsection{Invariance Algebra of $U(n)$}
The $U(n)$ algebra can be decomposed into $U(1) \times SU(n)$ where the generator of the central $U(1)$ is the diagonal unit matrix. For simplicity, let us consider $n=2$ first. A standard representation for the irreducible unitary representations of $SU(2)$ are the Pauli matrices, \begin{eqnarray}
\sigma_1=\left(\begin{array}{cc}0 &1 \\ 1 & 0 \end{array}\right)\qquad
\sigma_2=\left(\begin{array}{cc}0 &-i \\ i & 0 \end{array}\right)\qquad
\sigma_3=\left(\begin{array}{cc}1 & 0 \\ 0 & -1 \end{array}\right).\nonumber
\end{eqnarray}
Together with the identity matrix we have four generators of $U(2)$. Using Eq.~(\ref{eq:bosalgeb}) we can associate the following bilinears with them,
\begin{eqnarray}
\begin{array}{rcl}
I &\rightarrow & a_1^{\dagger}a_1 + a_2^{\dagger}a_2\\
\sigma_1 &\rightarrow & a_2^{\dagger}a_1 + a_1^{\dagger}a_2\\
\sigma_2 &\rightarrow & i(a_2^{\dagger}a_1 - a_1^{\dagger}a_2)\\
\sigma_3 &\rightarrow & a_1^{\dagger}a_1 - a_2^{\dagger}a_2
\end{array}\nonumber
\end{eqnarray}
Higher dimensional generalizations are well-known -- the standard representation for $SU(3)$ is given by the Gell-Mann matrices and a representation for $SU(n)$ with $n > 3$ by the generalized Pauli matrices.
We use these representations to read off the operator algebra. For $U(n)$ the operators $H$, $i L_{ij}$ and $T_{ij}$ are defined by,
\begin{eqnarray}
\begin{array}{rcll}
H&=&\sum_{i=1}^n a_i^{\dagger} a_i& \text{generator of the central } U(1)\\
L_{ij}&=&a_i^{\dagger} a_j - a_j^{\dagger} a_i & \text{generators of }SO(n) \subset SU(n)\\
T_{ij}&=&a_i^{\dagger} a_j + a_j^{\dagger} a_i-\frac{2}{n}\delta_{ij}\sum_{k=1}^na_k^{\dagger}a_k & \text{generators of the coset space } SU(n) / SO(n) 
\end{array}\label{eq:oscgen}
\end{eqnarray}
The operator $H$ is interpreted as energy; from $L_{ij}$ the complex unit was factored out so we are left with a real-valued second-order skew-symmetric tensor,
which is interpreted as the angular momentum of the harmonic oscillator. The second-order traceless symmetric tensor $T_{ij}$ is interpreted as the quadrupole operator of the harmonic oscillator. The operator expressions are valid for both the bosonic as well as the fermionic operator algebra.
\subsection{CSCO and Highest Weight Representation}
One choice of a Complete Set of Commuting Observables (CSCO) for the $U(n)$ quantum theory is given by the operators,
\begin{eqnarray}
N_i=a_i^{\dagger}a_i,
\end{eqnarray}
with $i=1,...,n$.
The operators $a_i^{\dagger}$ and $a_i$ act as raising and lowering operators for the operators $N_i$,
\begin{eqnarray}
\begin{array}{rcl}
\,[N_i,a_i^{\dagger}]&=&a_i^{\dagger},\\
\,[N_i,a_i]&=&-a_i.
\end{array}
\end{eqnarray}
For a normalized eigenstate $\lvert n_i\rangle$ we have,
\begin{eqnarray}
n_i = n_i\langle n_i \vert n_i\rangle=\langle n_i \vert a_i^{\dagger} a_i\vert n_i\rangle = \left| a \vert n_i\rangle \right|^2 \ge 0,
\end{eqnarray}
so we know that the eigenvalues $n_i$ are non-negative. In order for the bound to hold, there must be one state with eigenvalue zero for all $a_i$,
\begin{eqnarray}
a_i \vert 0\rangle =0,
\end{eqnarray}
otherwise repeated application of the lowering operators $a_i$ would invariably lead to states with negative eigenvalues, whose existence we just excluded. This is the highest weight state.
\subsection{Symmetry-breaking Interaction Terms}
Before discussing how to add interaction terms which preserve the symmetry, I briefly digress to arbitrary interaction terms. While arbitrary interaction terms generally break the $U(N)$ symmetry, they can nevertheless describe a system when embedded in a larger symmetry group.
For instance the harmonic oscillator as a sub-system of the non-relativistic quantum theory derived from Galilei-invariance can be described by the Schr\"odinger equation with arbitrary perturbations. In a suitable normalization the Hamiltonian of an harmonic oscillator with a $x^4$ interaction term is given by,
\begin{eqnarray}
H=H_f+gH_I=\frac{1}{2}(p^2+ x^2) +g x^4
\end{eqnarray}
With the identification,
\begin{eqnarray}
a_i=\frac{1}{\sqrt{2}}(x_i+ip_i)\qquad a_i^{\dagger}=\frac{1}{\sqrt{2}}(x_i-ip_i),
\end{eqnarray}
and its inverse,
\begin{eqnarray}
x_i=\frac{1}{\sqrt{2}}(a_i^{\dagger}+a_i) \qquad p_i=i\frac{1}{\sqrt{2}}(a_i^{\dagger}-a_{i}),
\label{eq:oscipx}
\end{eqnarray}
we obtain,
\begin{eqnarray}
H=\sum_i a_i^{\dagger}a_i +\frac{g}{4}\sum_i(a_i+a_i^{\dagger})^4.\nonumber
\end{eqnarray}
The system can be solved perturbatively by using only the properties of the operator algebra. This calculation has been performed for
example in~\cite{thesisXhu}.
\subsection{Interaction of the Quantum System with a Conserved Current}
I begin with the simplest symmetry-preserving interaction terms. When we add the primitive interaction terms to the generators Eq.~(\ref{eq:oscgen}) of the harmonic oscillator, we obtain,
\begin{eqnarray}
\begin{array}{ccl}
D^{-1}(\lambda)HD(\lambda)&=&\sum_{i=1}^n (a_i^{\dagger} a_i  +\lambda_i {a_i}^{\dagger}+\lambda_i^* a_i +  \lambda_i \lambda_i^*)\\
D^{-1}(\lambda)L_{ij}D(\lambda)&=&a_i^{\dagger} a_j - a_j^{\dagger} a_i -\lambda_i {a_j}^{\dagger} +\lambda_i^*a_j + \lambda_j a_i^{\dagger} -\lambda_j^*a_i -\lambda_i \lambda_j^* + \lambda_i^* \lambda_j\\
D^{-1}(\lambda)T_{ij}D(\lambda)&=&a_i^{\dagger} a_j + {a_j}^{\dagger} a_i+\lambda_i {a_j}^{\dagger} +\lambda_i^*a_j + \lambda_ja_i^{\dagger} +\lambda_j^*a_i +\lambda_i \lambda_j^* + \lambda_i^* \lambda_j\\
&&-\frac{2}{n}\delta_{ij}\sum_{k}(a_k^{\dagger}a_k+\lambda_k {a_k}^{\dagger}+\lambda_k^* a_k +  \lambda_k \lambda_k^*)
\end{array}
\end{eqnarray}
The above expressions illustrate some generic features of the primitive interactions. The free Hamiltonian is given in terms of bilinears. It picks up new terms linear in the creation and annihilation operators as well as a constant term (which may be infinite in the continuous case) which changes the normalization. The constant does not affect the dynamics and can be ignored for most purposes, but without it the algebra would not close. For the Hamiltonian derived above we can separate $\lambda_i \equiv g_ie^{i\pi\phi_i}$ into a real amplitude and a phase and obtain,
\begin{eqnarray}
H=\sum_{i=1}^n (a_i^{\dagger} a_i +g_i (e^{i\pi\phi_i} {a_i}^{\dagger}+e^{-i\pi\phi_i} a_i) +  g_i^2).
\end{eqnarray}
This Hamiltonian is familiar from quantum optics. It describes the interaction of a photon field with an external conserved current and is solved by the coherent states. When the source is time-dependent, the phase $\phi=\phi(t)$\ is also time-dependent and the photon field is interacting with a time-dependent current. For reference see~\cite{Glauber:1963fi,Glauber:1963tx} and for instance~\cite{Zhang:1999is}. In a quantum lattice the states are the phonons. Under the influence of the current the oscillators are uniformly displaced from their position on the lattice. Likewise their momentum can shift uniformly. This shift in the phase space, $x_i \rightarrow x_i+ Re(\lambda_i)$ and $p_i \rightarrow p_i+ Im(\lambda_i)$, is generated by the displacement operator.
\subsection{"Squeezed" States}
The previous section dealt with the lowest order interaction term. The next highest order interaction is generated by the transformation,
\begin{eqnarray}
\begin{array}{rcl}
a_j &\rightarrow& a_j\,+i\displaystyle g e^{-i\pi \phi}\left(e^{-i\pi \phi} a^{\dagger}_k +e^{i\pi \phi} a_k\right) \\
a_j^{\dagger} &\rightarrow& a_j^{\dagger}-i\displaystyle ge^{+i\pi \phi}\left(e^{-i\pi \phi} a^{\dagger}_k +e^{i\pi \phi} a_k\right)
\end{array}\qquad g \in \mathbb{R}\;\; \phi\in [0,2\pi].\nonumber
\end{eqnarray}
We will restrict to only one dimension and drop the subscripts on $a_j$ and $a_j^{\dagger}$. Under this transformation a free Hamiltonian $H_f=a a^{\dagger} + a^{\dagger} a$ turns into,
\begin{eqnarray}
\begin{array}{l}
H_{I}=(a a^{\dagger} + a^{\dagger} a)(1+2g)
+2\sqrt{2}ge^{-i\pi (\phi-\frac{1}{4})}a^{\dagger 2}
+2\sqrt{2} ge^{i\pi (\phi-\frac{1}{4})}a^{2}.
\end{array}\label{eq:squeezed}
\end{eqnarray}
The interaction term describes the exchange of two quanta. The analogy of the system is a child on a swing who interacts twice in one oscillation period by kicking his legs backward respectively forward at the highest points in the front and in the back.
The solution to the interacting Hamiltonian can be obtained from the solutions of the free theory by acting with the corresponding operator on the eigenstates.
\subsection{Transformations under the Invariance Algebra}
A transformation generated by 
$a^{\dagger}_iM_{ij}a_j$ where $M_{ij}$ is anti-Hermitian
has been discussed below Eq.~(\ref{eq:bilineartrans}). For
the harmonic oscillator any such operator is part of the invariance algebra and as such leaves the physics invariant. For instance for real $M_{ij}$ the
operator $a^{\dagger}_iM_{ij}a_j$ is proportional to the generators of angular momentum $L_{ij}$ and therefore generates a rotation under which the harmonic oscillator is invariant. The rotation matrix is $R=e^{a^{\dagger}_iM_{ij}a_j}$ and is orthogonal since $M_{ij}$ is anti-Hermitian. Therefore the Hamiltonian is invariant under this transformation,
\begin{eqnarray}
a_{i}^{\dagger}a_i\longrightarrow a_{i}^{\dagger}R_{ij}R_{ji}a_i=a_{i}^{\dagger}a_i.
\end{eqnarray}
\subsection{The Point Particle in a Magnetic Field}
There is no reason to restrict oneself exclusively to the diagonal $U(1)$ representation to define a Hamiltonian. In particular, one may add other generators of the invariance algebra to it. To illustrate this and because of the similarity to the bosonic string it will be shown in this subsection that $H_0(\omega)-\omega L_3$ corresponds to the point particle under the influence of a magnetic field. Consider a  magnetic field in the $x_3$-direction,
\begin{eqnarray}
\vec{B}(\vec{x})=(0,0,B_3(\vec{x})).
\end{eqnarray}
We associate with it the vector potential,
\begin{eqnarray}
\vec{A}(\vec{x})=(A_1(x_1,x_2),A_2(x_1,x_2),0),
\end{eqnarray}
since,
\begin{eqnarray}
\vec{B}=\nabla \times \vec{A}=(0,0,\partial_1A_2(x_1,x_2)-\partial_2A_1(x_1,x_2)).
\end{eqnarray}
For a constant magnetic field $B_{3}(\vec{x})=B$ a possible solution is
$\vec{A}=\frac{B}{2}(x_2,x_1,0)$ which is at the same time
in the Coulomb gauge as well as in Poincar\'e gauge.
The Hamiltonian is,
\begin{eqnarray}
H_{B}=\frac{1}{2m}(\vec{p}-\frac{q}{c}\vec{A}(\vec{x}))^2
=\frac{1}{2m}((p_1-\frac{q}{c}A_1(\vec{x}))^2
+(p_2-\frac{q}{c}A_2(\vec{x}))^2+p_3^2).
\end{eqnarray}
Since the wave function can be factorized and the third dimension describes a free particle, the problem in the two non-trivial dimensions reduces to,
\begin{eqnarray}
\begin{array}{rcl}
H_{B}&=&\displaystyle\frac{1}{2m}((p_1+\frac{Bq}{2c}x_2)^2
+(p_2-\frac{Bq}{2c}x_{1})^2)\\
&=&\displaystyle\frac{p_1^2+p_2^2}{2m}+\frac{B^2q^2}{8mc^2}(x_1^2+x_2^2)
-\frac{Bq}{2mc}(x_{1}p_2-x_2p_1).
\end{array}
\end{eqnarray}
This Hamiltonian can be decomposed into the Hamiltonian of the harmonic oscillator
$H_{0}(\omega)=\frac{1}{2m}\vec{p}^2+\frac{1}{2} m\omega^2\vec{x}^2$ and the generator $L_3=x_{1}p_2-x_{2}p_1$ of angular momentum. One has,
\begin{eqnarray}
H_{B}=H_{ho}(\omega_B)-\omega_BL_3 \qquad \omega_B=\frac{Bq}{2mc}.
\end{eqnarray}
We can further express these operators in terms of the eigenstates of the angular momentum operator,
\begin{eqnarray}
a=\frac{1}{\sqrt{2}}(a_1+ia_2)\qquad 
\tilde{a}=\frac{1}{\sqrt{2}}(a_1-ia_2),
\end{eqnarray}
so that,
\begin{eqnarray}
\begin{array}{l}
H_{ho}(\omega)=\hbar\omega
 (a^\dagger a-\tilde{a}^\dagger \tilde{a}) \qquad  L_3=-\hbar
 (a^\dagger a-\tilde{a}^\dagger \tilde{a}).\\
 \end{array}
\end{eqnarray}
The states can be written $|N,\bar{N}\rangle\propto (a^{\dagger})^N (\tilde{a}^{\dagger})^{\bar{N}}|0\rangle$ and the eigenvalue equations are,
\begin{eqnarray}
\begin{array}{rcl}
H|N,\bar{N}\rangle&=&\hbar \omega (N+\bar{N}+1)|N,\bar{N}\rangle\\
L_{3}|N,\bar{N}\rangle&=&\hbar (N-\bar{N})|N,\bar{N}\rangle.
\end{array}
\end{eqnarray}
The states can be labeled by the linearly independent combinations $N+\bar{N}$ and $N-\bar{N}$. 
\subsection{The Supersymmetric Harmonic Oscillator}
Introducing interactions to a supersymmetric system is not much different from working with a non-supersymmetric system. The algebra is enlarged, which can make things more unwieldy, but there are no principal difficulties. In a supersymmetric system with $U(N)$ symmetry the algebra is further extended
by the algebra of the supercharges:
\begin{eqnarray}
\{Q,Q^{\dagger}\}=2H\qquad \{Q,Q\}=0\qquad \{Q^{\dagger},Q^{\dagger}\}=0\qquad [Q,H]=0
\end{eqnarray}
Here $H$ is a sum of the bosonic and the fermionic oscillator representation of the identity,
\begin{eqnarray}
H=H_{bos} + H_{fer} = a^{\dagger}a + b^{\dagger}b. 
\end{eqnarray}
The algebra is satisfied by the supercharge,
\begin{eqnarray}
Q=i a b^{\dagger}.
\end{eqnarray}
All one needs to do is apply the same unitary transformation to the supercharge $Q$. Under the action of the displacement operator one has for instance,
\begin{eqnarray}
D(\lambda)QD(\lambda)^{-1}&=&iab^{\dagger}-i\lambda b^{\dagger} \qquad \lambda \in \mathbb{C}.
\end{eqnarray}

%% file: StringBgr_examplePoincare.tex
\section{Translation, Poincar\'e and Conformal Group}
As mentioned in the introduction, operators in
non-relativistic quantum mechanics depend on continuous parameters. However,
discrete operator representations also exist.
In particular, the algebra of string theory
contains such a discrete representations as a subalgebra.  While it is possible to work with continuous representations~\cite{klink2015relativity}, it may be impossible to express the operators of the invariance algebra in terms of bilinears. This paper restricts itself to discrete operator representations only. This section will discuss  the translation group, the Lorentz group, the Poincar\'e group and the conformal group. 
\subsection{Translation Group}
\subsubsection{Translation Algebra in Terms of Bilinears}
For the algebra $[\hat{P}^{\mu},\hat{P}^{\nu}]=0$ of the translation group the complete set of commuting observables consist of the generators $\hat{P}^{\mu}$.
That means we can label the states by their eigenvalues.
It is easy to see that for any self-adjoint operator $\hat{P}$ the operator $U(c)=e^{i\hat{P}c}$ is a
unitary irreducible representation of the translation group. Unitarity requires $\hat{P}$ to be Hermitian.
Hermitian operators have real eigenvalues. Basis of $U(x)$ is the
eigenvector $|p\rangle$\ of $\hat{P}$:
\begin{eqnarray}
\hat{P}|p\rangle=p |p\rangle \qquad U(x)|p\rangle = e^{-ipx}|p\rangle.
\end{eqnarray}
The generalization to higher dimensions is obvious.
\subsection{Alternative Representations of the Translation Group}
The translation algebra appears as part of other algebras. In particular, the Poincar\'e group is a semi-direct product of the translation group and the Lorentz group. For that reason it can be useful to have
an operator representation other than the one above, which are\ better compatible with the embedding Poincar\'e algebra.
Consider the following matrix representation of the translation group of $\mathbb{R}^3$:
\begin{eqnarray}
\rho(c_{i})=\left(
\begin{array}{cccc}
1 & 0 & 0 & c_1\\
0 & 1 & 0 & c_2\\
0 & 0 & 1 & c_3\\
0 & 0 & 0 & 1
\end{array}
\right).
\end{eqnarray}
The representation satisfies $\rho(c_{i})\rho(d_i)=\rho(c_i+d_i)$ and $\rho(c_{i})^{-1}=\rho(-c_{i})$.
It is generated by,
\begin{eqnarray}
\left(
\begin{array}{cccc}
0 & 0 & 0 & c_{1}\\
0 & 0 & 0 & c_{2}\\
0 & 0 & 0 & c_{3}\\
0 & 0 & 0 & 0
\end{array}
\right),
\end{eqnarray}
since,
\begin{eqnarray}
\exp\left(
\begin{array}{cccc}
0 & 0 & 0 & c_{1}\\
0 & 0 & 0 & c_{2}\\
0 & 0 & 0 & c_{3}\\
0 & 0 & 0 & 0
\end{array}
\right)=\left(
\begin{array}{cccc}
1 & 0 & 0 & c_1\\
0 & 1 & 0 & c_2\\
0 & 0 & 1 & c_3\\
0 & 0 & 0 & 1
\end{array}
\right).
\end{eqnarray}
We can construct a bilinear operator representation out of this representation whose generators are,
\begin{eqnarray}
 a_1^{\dagger}a_4,\;
 a_2^{\dagger}a_4,\;
 a_3^{\dagger}a_4.
\end{eqnarray}
Since $a_4$ plays no further role it may as well be dropped.
The generators $a_1^{\dagger}, a_2^{\dagger}, a_3^{\dagger}$ are however not self-adjoint. For unitary representations $U(c)=e^{i\hat{P}c}$ we need $\hat{P}$ to be self-adjoint. We can define $\hat{P}_i(\theta)=a^{\dagger}_{i}e^{-i\theta}+a_{i}e^{i\theta}$
where $\theta$ is an arbitrary phase and the creation and annihilation operators satisfy the Heisenberg-Weyl algebra. Specifically
we can choose $\theta=90^o$ so that,
\begin{eqnarray}
\hat{P_{i}}=i(a_{i}^{\dagger}-a_i) \qquad U(x)=e^{i\hat{P}_{i}x_{i}}.
\end{eqnarray}
The signs in the exponent of the unitary operator $U(x)$ are independent for each direction and purely conventional. They can be made compatible with the Minkowski metric if desired. The form above is how the translation operators appears in the Poincar\'e algebra. The general rule has been that creation and annihilation operators appear as bilinears when they generate symmetry transformations and appear alone when they generate interactions. In this special case they appear alone but nevertheless generate symmetry transformations. 
\subsection{Lorentz and Poincar\'e algebra}
A good review on the unitary irreducible representations of the Lorentz and Poincar\' e group is~\cite{Bekaert:2006py}. The Lie algebra of the proper orthochronous Lorentz group $SO(1,D-1)^{+}$ can be written in the form,
\begin{eqnarray}
\begin{array}{rcl}
i\left[M_{\mu\nu},M_{\rho\sigma}\right]&=&
\left(\eta_{\nu\rho}M_{\mu\sigma}+\eta_{\mu\sigma}M_{\nu\rho}-\eta_{\mu\rho}M_{\nu\sigma}-\eta_{\nu\sigma}M_{\mu\rho}\right)
\end{array}\label{eq:lorentzalg}
\end{eqnarray}
where Greek indices take values $0$ to $D-1$. For $D \ge 4$ the centrally extended Lorentz groups are their double cover $\text{Spin}(1,D-1)$. The simplest non-trivial representation of the Lorentz algebra is its fundamental representation. We now want to express the generators of the fundamental representation in terms of bilinears of creation and annihilation operators. It is convenient to slightly modify the canonical commutation relationships for compatibility with the Minkowski metric $\eta_{\mu\nu}$ and use,
\begin{eqnarray}
\; [a_\mu,a^{\dagger}_\nu]=\eta_{\mu\nu}
\qquad [a_\mu,a_\nu]=0
\qquad [a_\mu^{\dagger},a^{\dagger}_\nu]=0.
\end{eqnarray}
Then the generators corresponding to the fundamental representations are,
\begin{eqnarray}
\begin{array}{rcl}
M_{\mu\nu}&=&-i(a_{\mu}^{\dagger} a_{\nu} - a_{\nu}^{\dagger} a_{\mu}) \\
\end{array}\qquad \mu\ne \nu\qquad \mu,\nu=0,...,D-1.
\end{eqnarray}
The Poincar\'e algebra is a further extension of the Lorentz algebra and is given by,
\begin{eqnarray}
\begin{array}{rcl}
i\left[M_{\mu\nu},M_{\rho\sigma}\right]&=&\left(\eta_{\nu\rho}M_{\mu\sigma}+\eta_{\mu\sigma}M_{\nu\rho}-\eta_{\mu\rho}M_{\nu\sigma}-\eta_{\nu\sigma}M_{\mu\rho}\right)\\
i\left[P_{\rho},M_{\mu\nu}\right]&=&\left(\eta_{\rho\mu} P_{\nu}-\eta_{\rho\nu}P_{\mu}\right)\\
\left[P_{\mu},P_{\nu}\right]&=&0
\end{array}\label{eq:poincarealg}
\end{eqnarray}
The general way to proceed as proposed in this paper is to select a matrix representation of the full algebra and express it in terms of bilinears. Here an easier way exists since the remaining relations of Eq.~(\ref{eq:poincarealg}) are satisfied when $P_\mu$ is an arbitrary linear combination of $a_\mu$ and $a^{\dagger}_\mu$.
If one wishes to maintain a bilinear structure, one could introduce an operator $c$ to form the expressions $c^{\dagger}a_\mu $ respectively $a^{\dagger}_\mu c$. Evaluating the algebra shows that it must satisfy
$[c,c^{\dagger}]=0$, and therefore $c=c^{\dagger}$ is  the central element.
One way to understand why the second operator in the bilinear combination is trivial, is by deriving the Poincar\'e algebra from the Lorentz algebra via an In\"on\"u-Wigner contraction~\cite{InonuWigner:1953}. In this process one singles out a dimension $N$ of the Lorentz algebra,
\begin{eqnarray}
\begin{array}{rcl}
i\left[M_{\mu\nu},M_{\rho\sigma}\right]&=&\left(\eta_{\nu\rho}M_{\mu\sigma}+\eta_{\mu\sigma}M_{\nu\rho}-\eta_{\mu\rho}M_{\nu\sigma}-\eta_{\nu\sigma}M_{\mu\rho}\right)\\
i\left[M_{\rho N},M_{\mu\nu}\right]&=&\left(\eta_{\rho\mu} M_{\nu N}-\eta_{\rho\nu}M_{\mu N}\right)\\
i\left[M_{\mu N},M_{\nu N}\right]&=&M_{\mu\nu}\\
\end{array}
\end{eqnarray}
 then defines $M_{\mu N}= R P_{\mu}$ and  increases the orbit of the action to infinite size by taking the limit $R \rightarrow \infty$ with $P_\mu$ fixed. In the limit one recovers the Poincar\'e algebra of Eq.~(\ref{eq:poincarealg}).
In the process, the $N$-th dimension vanishes and the associated operators $a^{\dagger}_N$ and $a_N$ become trivial. For completeness, we note the matrix representation of the 
Poincar\'e algebra in two dimensions. The representation consists of the boost generator, the generator of time translations and the generator of space translations:
\begin{eqnarray}
\left(
\begin{array}{cccc}
0 & -1 & 0\\
-1 & 0 & 0\\
0 & 0 & 0
\end{array}
\right),
\left(
\begin{array}{cccc}
0 & 0 & 1\\
0 & 0 & 0\\
0 & 0 & 0
\end{array}
\right),
\left(
\begin{array}{cccc}
0 & 0 & 0\\
0 & 0 & 1\\
0 & 0 & 0
\end{array}
\right)
\end{eqnarray}
The resulting bilinear representation is,
\begin{eqnarray}
M_{01}=-i(a_0^{\dagger}a_1+a_1^{\dagger}a_0)
\qquad
P_0 = ia_0^{\dagger}a_3 \qquad P_1 = ia_1^{\dagger}a_3.
\end{eqnarray}
Since $a_3$ commutes with all other operators which appear,
it plays the role of a $c$-number and can be dropped.
\subsection{Position-Momentum Duality}
The Lorentz generators can be written as,
\begin{eqnarray}
\hat{M}_{\mu\nu}=\frac{1}{2}(\hat{X}_\mu \hat{P}_\nu-\hat{X}_\nu \hat{P}_\mu).
\end{eqnarray}
Comparing with the operator expression $\hat{M}_{\mu\nu}=-i(a_\mu^{\dagger} a_\nu - a_\nu^{\dagger} a_\mu)$ one could be tempted to identify for instance,
\begin{eqnarray}
\hat{X}_\mu \stackrel{?}{=}a_\mu^{\dagger}
\qquad
\hat{P}_\mu \stackrel{?}{=}-ia_\mu.
\end{eqnarray}
But we are looking for a choice where $\hat{X}^{\mu}$ and $\hat{P}^{\mu}$\ are self-adjoint. The rotation,
\begin{eqnarray}
\begin{array}{rcl}
a_{\mu}^{\dagger} &\rightarrow& \cos(\phi)a_{\mu}^{\dagger} +\sin(\phi)a_{\mu}\\
a_i &\rightarrow& -\sin(\phi)a_{\mu}^{\dagger}+\cos(\phi)a_{\mu},
\end{array} \label{eq:rotaadagger}
\end{eqnarray}
leaves $[a_{\mu},a_{\nu}^{\dagger}]=\eta_{\mu\nu}$ as well as the generators of the form $a_{\mu}^{\dagger}a_{\nu}-a_{\mu}^{\dagger}a_{\nu}$ invariant. The angle $\phi$ is independent of ${\mu}$. This rotation does not preserve the property that $a_{\mu}$ and $a_{\mu}^{\dagger}$ are
adjoint of each other. To find self-adjoint operators 
$\hat{X}_{\mu}$ and $\hat{P}_{\mu}$ this is useful, but
it must be remembered that this is not a symmetry of the physical theory. Using this rotation together with an overall rescaling we can identify the self-adjoint operators familiar from the harmonic oscillator:
\begin{eqnarray}
\hat{X}_{\mu}:=\frac{1}{\sqrt{2}}(a_{\mu}^{\dagger}+a_{\mu}) \qquad 
\hat{P}_{\mu}:= \frac{i}{\sqrt{2}}(a_{\mu}^{\dagger}-a_{\mu}).
\label{eq:px}
\end{eqnarray}
It is easy to check that with this definition the Lorentz generator is still $\hat{M}_{\mu\nu}=\frac{1}{2}(\hat{X}_\mu \hat{P}_\nu-\hat{X}_\nu \hat{P}_\mu)
=-i(a_{\mu}^{\dagger} a_\nu - a_{\nu}^{\dagger} a_{\mu})$ and that the canonical commutation relations,
\begin{eqnarray}
[\hat{X}_{\mu},\hat{P}_{\nu}]=i\eta_{\mu\nu},
\end{eqnarray}
are automatically satisfied. Based on Eq.~(\ref{eq:px}) the operators $\hat{X}_{\mu}$ and $\hat{P}_{\mu}$ can be regarded as the real and the imaginary component in a complex space. The algebra -- and consequently the entire physics of the theory -- is invariant under rotations in this complex plane $z \rightarrow z e^{ i\theta}$:
\begin{eqnarray}
\begin{array}{ccc}
e^{i\theta \hat{N_{k}}}a_ke^{-i\theta \hat{N_{k}}}&=&e^{-i\theta}a_{k},\\
e^{i\theta \hat{N_k}}a^{\dagger}_ke^{-i\theta \hat{N_k}}&=&e^{i\theta}a^{\dagger}_{k}.
\end{array}\label{eq:numop}
\end{eqnarray}
Here $\hat{N}_k=a_k^{\dagger}a_k$ is the number operator.
In particular, a rotation by $\theta=\pi/2$ maps $\hat{X}_{\mu} \rightarrow \hat{P}_{\mu}$ and $\hat{P}_{\mu} \rightarrow -\hat{X}_{\mu}$.
This symmetry is the well-known space-momentum duality. 
The core of quantum theory is its  use of projective representations. While the initial symmetry may appear to be real-valued, the use of projective representations invariably introduces a complex space. The distinction between the real and imaginary component is blurred by the identification of its elements up to a phase.
This complex space is the phase space.
In the derivation of the free theory at the beginning of this paper the symmetry was not further specified. I would conjecture that it should take the form,
\begin{eqnarray}
(x,t) \mapsto f(x,t;p,E),
\end{eqnarray}
that is, act on the phase space.
\subsection{Transformation Properties under the Poincar\'e Group}
In the fundamental representation the generators of the Lorentz group are nothing but the $SO(1,D-1)^+$ matrices themselves, that is both $\Lambda^{\rho}{}_{\sigma}=\exp(-\frac{i}{2}\omega_{\mu\nu}M^{\mu\nu})^{\rho}{}_{\sigma}$ and $\Lambda^{\rho}{}_{\sigma}=(e^{\omega})^{\rho}{}_{\sigma}$ are
the usual Lorentz transformation matrices. The explicit form of the generators is $(M^{\mu\nu})^{\rho}{}_{\sigma}=i(\eta^{\mu\rho}\delta^{\nu}{}_{\rho}-\eta^{\nu\rho}\delta^{\mu}{}_{\sigma})$. From the corresponding bilinear operators we derive the transformation property of the creation and annihilation operators from our templates of transformation and find,
\begin{eqnarray}
\begin{array}{rcl}
e^{-\frac{i}{2}\omega_{\mu\nu}M^{\mu\nu}}a_{\kappa}e^{\frac{i}{2}\omega_{\mu\nu}M^{\mu\nu}}&=&(e^{-\omega})_{\kappa}{}^{\lambda}a_{\lambda}=(\Lambda^{-1})^{\lambda}{}_{\kappa}a_{\lambda}\\
e^{-\frac{i}{2}\omega_{\mu\nu}M^{\mu\nu}}a_{\kappa}^{\dagger}e^{\frac{i}{2}\omega_{\mu\nu}M^{\mu\nu}}&=&(e^{-\omega})_{\kappa}{}^{\lambda}a^{\dagger}_{\lambda}=(\Lambda^{-1})^{\lambda}{}_{\kappa}a^{\dagger}_{\lambda}
\end{array}
\end{eqnarray}
since $(\Lambda^{-1})^{\lambda}{}_{\kappa}=\Lambda_{\kappa}{}^{\lambda}$.
Using $\hat{P}_{\mu}= i(a_{\mu}^{\dagger}-a_{\mu})$ we recover the familiar Lorentz transformation law,
\begin{eqnarray}
U(\Lambda)\hat{P}_{\kappa}U(\Lambda)^{-1}=(\Lambda^{\lambda}{}_{\kappa})^{-1}\hat{P}_{\lambda}.
\end{eqnarray}
Elements of the Poincar\'e group are parametrized by the real variables $c^{\mu}$ and $\omega^{\mu\nu}=-\omega^{\nu\mu}$ and are  obtained by the exponential map,
\begin{eqnarray}
g(c^{\mu},\omega^{\mu\nu})=\exp\left\{i(-c^{\mu}P_{\mu}+\frac{1}{2}\omega^{\mu\nu}M_{\mu\nu})\right\}.
\end{eqnarray}
\subsection{Conformal Algebra}
The Poincar\'e algebra can be further extended to a conformally invariant algebra. Again we want to find a representation in terms of bilinear
operators. It would be possible to proceed in the same manner as before and find matrix representations
of the algebra from which one then sets up the bilinears. Here we go an alternative way.
Given a symmetry, the starting point is usually a set of
finite transformations. It is straightforward to derive the
infinitesimal transformations from them. In case of
the conformal algebra they are~\cite{DiFrancescoCFT},
\begin{eqnarray}
\begin{array}{ccl}
M_{\mu\nu}&=&-i(x_{\mu}\partial_{\nu}-x_{\nu}\partial_{\mu})\\
P_{\mu}&=&-i\partial_{\mu}\\
D&=&-ix^{\mu}\partial_{\mu}\\
K_{\mu}&=&-i(2x_{\mu}x^{\nu}\partial_{\nu}-x^{\nu}x_{\nu}\partial_{\mu})\\
\end{array}
\label{eq:confalgd}
\end{eqnarray}
They satisfy the algebra,
\begin{eqnarray}
\begin{array}{rcl}
i\left[M_{\mu\nu},M_{\rho\sigma}\right]&=&\left(\eta_{\nu\rho}M_{\mu\sigma}+\eta_{\mu\sigma}M_{\nu\rho}-\eta_{\mu\rho}M_{\nu\sigma}-\eta_{\nu\sigma}M_{\mu\rho}\right )\\
i\left[P_{\rho},M_{\mu\nu}\right]&=&\eta_{\rho\mu} P_{\nu}-\eta_{\rho\nu}P_{\mu}\\
i\left[K_{\rho},M_{\mu\nu}\right]&=&\eta_{\rho\mu} K_{\nu}-\eta_{\rho\nu}K_{\mu}\\ 
\left[P_{\mu},P_{\nu}\right]&=&0\qquad \left[K_{\mu},K_{\nu}\right]=0\\
\left[D,P_{\mu}\right]&=&iP_{\mu}\qquad \left[D,K_{\mu}\right]=-iK_{\mu} \qquad \left[D,M_{\mu\nu}\right]=0\\
\left[K_{\mu},P_{\nu}\right]&=&2i(\eta_{\mu\nu}D+M_{\mu\nu})
\end{array}\label{eq:conformalalg}
\end{eqnarray}
To find a representation in terms of bilinears one can make use of the isomorphism between $\left[\partial_i,x_j\right]=\delta_{ij}$
and $[a_i,a_j^{\dagger}]=\delta_{ij}$.
From Eq.~(\ref{eq:confalgd}) one can directly read off,
\begin{eqnarray}
\begin{array}{rcl}
M_{\mu\nu}&=&-i(a_{\mu}^{\dagger} a_{\nu} - a_{\nu}^{\dagger} a_{\mu})\\
P_{\mu}&=&-ia_{\mu}\\
K_{\mu}&=&-i(2a^{\dagger}_{\mu} a^{\nu\dagger}a_{\nu}-a^{\dagger\nu}a^{\dagger}_{\nu} a_{\mu})\\
 D&=&-ia^{\nu\dagger}a_{\nu}
\end{array}
\end{eqnarray}
As discussed earlier, this mapping ignores that $x_{i}$ and $\partial_{i}$ are not adjoint
of each therefore $a_i$ and $a_i^{\dagger}$ above are also not adjoint pairs, contrary to what the notation suggests. Using the isomorphism between $\left[\partial_i,x_j\right]=\delta_{ij}$
and $[\frac{1}{\sqrt{2}}(a_i^{\dagger}-a_i),\frac{1}{\sqrt{2}}(a_i^{\dagger}+a_i)]=\delta_{ij}$ would remedy that. Then the momentum matches with the harmonic oscillator momentum 
$P_{\mu}=\frac{i}{\sqrt{2}}(a_i^{\dagger}-a_i)$ and the dilatation matches with the generator of the squeezing operator 
$D=-\frac{i}{2}(a^{\nu\dagger}a_{\nu}^{\dagger}-a^{\nu}a_{\nu}+d-2)$.

%% file: StringBgr_exampleString.tex
\section{String Theory}
\subsection{Two-dimensional conformal symmetry}
The world-sheet description of the (super-)string is based on the
two-dimensional (super-) conformal group. A variety of different treatments can be found in the literature, such as in~\cite{DiFrancescoCFT} but no single reference follows our approach completely. For the benefit of readers not intimately familiar with string theory and to further clarify the symmetry based derivation, it is helpful to collect some
results here. The reader can skip or skim this subsection.
\subsubsection{Symmetry Group in Complex Coordinates}
Conformal symmetry is rather simple in complex coordinates. Any holomorphic map of the complex plane into itself,
\begin{eqnarray}
z \rightarrow w(z),
\end{eqnarray}
is a conformal mapping. The scaling factor and the rotation angle become manifest in the differential:
\begin{eqnarray}
dw = \frac{dw}{dz}dz=\left|\frac{dw}{dz}\right|e^{i\text{arg}\left(\frac{dw}{dz}\right)}dz
\end{eqnarray}
Anti-holomorphic transformations are treated in an analogous manner.
\subsubsection{Conformal Generators}
An infinitesimal (anti-)holomorphic mapping admits a Laurent expansion from which the conformal generators can be read off:
\begin{eqnarray}
\begin{array}{rcl}
l_n&=&-z^{n+1}\partial_z\\
\bar{l}_n&=&-\bar{z}^{n+1}\partial_{\bar{z}}
\end{array} \qquad \text{ for }n \in \mathbb{Z}
\end{eqnarray}
\subsubsection{Conformal Algebra}
The generator satisfy the conformal algebra,
\begin{eqnarray}
\begin{array}{rcl}
\left[l_n,l_m\right]&=&(n-m)l_{n+m}\\
\left[\bar l_n,\bar l_m\right]&=&(n-m)\bar{l}_{n+m}\\
\left[l_n,\bar l_m\right]&=&0
\end{array}
\end{eqnarray}
which is referred to as Witt algebra. The transformations considered so far are local transformations. The subalgebra of global conformation transformations is generated by $l_{-1}$, $l_0$ and $l_1$. Specifically, translations on the complex plane are generated by $l_{-1}=-\partial_z$, scale transformations and rotations by $l_0=-z\partial_z$ and special conformal transformations by $l_1=-z^2 \partial_z$.
\subsubsection{Central Extension}
The central extension of the Witt algebra is the Virasoro algebra and derivations of the central extension of the Witt algbra can be found for instance in~\cite{deKerf:1997ga,kac2013bombay,khesin2008geometry}.
It turns out that the only non-vanishing element is $\omega_{n,-n}$ and the solution space of our 2-cocycles is given by $\omega_{n,m}=(\lambda_1 n+\lambda_2 n^{3})\delta_{m,-n}$ where $\lambda_1,\lambda_2\in\mathbbm{C}$. 
The basis element $\omega_{n,-n}=n$
is an exact element and any term proportional to it can be absorbed in the algebra. Therefore one could set $\lambda_1=0$. By rescaling the operators one could also set $\lambda_2=1$ so that $\omega_{n,m}=n^3\delta_{m,-n}$ is the unique solution. The conventional choice for the extended algebra is however,
\begin{eqnarray}
\left[L_n,L_m\right]&=&(n-m)L_{n+m}+\frac{c}{12}(m^{3}-m)\delta_{m,-n}
\end{eqnarray}
Up to equivalences and rescaling this is the unique non-trivial extension of the Witt algebra. These equivalences are the simplest of the automorphisms which give rise to interactions.
\subsubsection{The Invariance Algebra in the Oscillator Representation}
The generators of the Virasoro algebra can be expressed in terms of the generators of the operator algebra of the harmonic oscillator. The oscillator states will be interpreted as excitations of the string.
The algebra is infinite dimensional and we define $\alpha_{-n}=\alpha_n^{\dagger}$.
It is convenient to use the normalization $\alpha_n \rightarrow \sqrt{n}\alpha_n$ for
the operators with $n \ne 0$ so that,
\begin{eqnarray}
\left[\alpha_m, \alpha_n \right]&=&m \delta_{m+n,0} \qquad m,n \in \mathbbm{Z}.
\label{eq:modes}
\end{eqnarray}
The normal ordering operator is defined as,
\begin{eqnarray}
:\alpha_i \alpha_j:=\left\{
\begin{array}{l}
\alpha_i \alpha_j \text{ for } i \le j\\
\alpha_j \alpha_i \text{ for } i > j
\end{array}
\right.
\end{eqnarray}
and the generator,
\begin{eqnarray}
L_n=\frac{1}{2}\sum_{j\in \mathbb{Z}}:\alpha_{j}\alpha_{n-j}:
\end{eqnarray}
It can be verified that the $L_n$ satisfy the Virasoro algebra with central charge $c=1$.
\subsubsection{Complete Set of Commuting Observables}
The maximal set of generators which commute with all the generators of
the algebra is given by $L_0$ and the central charge $c$. Here the choice of $L_0$ was arbitrary since any one generator could have been chosen to
be diagonalized in addition to $c$. The eigenvalues of $L_0$ and $c$ will be called $h$ and $c$ respectively, so we can label states by $\lvert h,c\rangle$.
\subsubsection{Highest Weight Representation}
Analogous to the angular momentum in quantum mechanics, $L_{-n}$ and $L_{n}$ are raising and lowering operators for the eigenvalues of $L_0$,
\begin{eqnarray}
\left[L_0,L_{\pm n}\right]=\mp L_{\pm n}.
\end{eqnarray}
From the algebra we see that $L_n$ with $n>0$ reduces the eigenvalue of $L_0$ by $n$,
\begin{eqnarray}
L_0 L_n \lvert \psi,c\rangle = \left[L_n L_0 - n L_n\right]\lvert \psi,c\rangle =(\psi-n) L_n \lvert \psi,c\rangle
\end{eqnarray}
The highest weight representation (although the better term here would be lowest-weight representation) is a representation with a state containing the lowest eigenvalue of $L_0$. If $\lvert h,c\rangle$ is a highest weight state, it must be annihilated by all $L_n$ with $n>0$, otherwise their action would reduce the eigenvalue even further, which would be contradictory to $\lvert h,c\rangle$ being a highest weight state. The physical states of the theory therefore satisfy,
\begin{eqnarray}
(L_0-h) \lvert h,c \rangle =0 \qquad L_n \lvert h,c \rangle =0 \text{ for }n>0.
\end{eqnarray}
In the language of CFT, these highest-weight states are also referred to as {\it primary} states. The first equation will be interpreted as the mass-shell condition.
The $L_{-n}$ with $n>0$ do not annihilate the highest weight state and their action can be used to generate descendant states.
\subsubsection{The Hilbert Space}
New states, the so-called descendant states, are obtained by applying the raising operators in all possible ways on highest weight states:
\begin{eqnarray}
L_{-k_1} L_{-k_2}...L_{-k_l}\lvert h,c\rangle \qquad 1 \le k_1 \le ... \le k_l.
\end{eqnarray}
By convention, the raising operators are arranged in increasing order of $k_i$. Their $L_0$ eigenvalue is $h+\sum_i k_i$. By the Poincar\'e-Birkhoff-Witt theorem this procedure generates the entire Hilbert space.
From the algebra it can be seen that the states can also be genßerated iteratively from a highest weight state:
\begin{eqnarray}
\begin{array}{c}
\lvert h,c\rangle\\
L_{-1} \lvert h,c\rangle\\
L_{-2} \lvert h,c\rangle, \;\; L_{-1} L_{-1} \lvert h,c\rangle\\
L_{-3} \lvert h,c\rangle, \;\; L_{-1} L_{-2} \lvert h,c\rangle,\; \; L_{-1}L_{-1} L_{-1} \lvert h,c\rangle\\
\vdots
\end{array}
\end{eqnarray}
The set of these states is called a Verma module $V(h,c)$. States in one Verma module are not necessarily linearly independent. If a linear combination of states of a Verma module vanishes, this linear combination is referred to as null state.
States in the same Verma module but at different levels are orthogonal to each other. If we chose a basis of orthogonal highest-weight states, that is $\langle h',c\vert h, c\rangle = 0$ for all $h \ne h'$, then states in different Verma modules are also orthogonal to each other.
\subsubsection{The Vacuum State}
The vacuum state $\lvert 0 \rangle$ is the most symmetric state of the theory which  means it should be annihilated by the largest possible number of conserved charges. In particular it must be invariant under the global conformal transformations $L_{-1}$, $L_0$ and $L_1$. The highest weight state $|h,c\rangle$ is already annihilated by all $L_n$ with $n>0$, so we turn our attention to the remaining operators. We want to find those $n\ge 0$ for which $L_{-n}\lvert 0 \rangle =0$. For them holds,
\begin{eqnarray}
0 =L_{-n} \lvert 0 \rangle = \left[L_n,L_{-n}\right] \lvert 0\rangle =\left(2n L_0 +\frac{1}{12}(n^{3}-n)c\right)\lvert 0 \rangle \label{eq:vac}  
\end{eqnarray}
The equation is satisfied for $n=0$, i.e. $L_0 \lvert 0 \rangle=0$ with eigenvalue $h=0$. Due to $L_0 \lvert 0 \rangle=0$, Eq.~(\ref{eq:vac}) is also satisfied for $n=1$.
\subsubsection{Unitary Representations}
For a unitary representation the inner product must be positive definite, otherwise $L_n^\dagger = L_{-n}$ does not hold. Using the algebra and the condition for physical states we find,
\begin{eqnarray}
\langle \phi\vert L_n L_{-n} \vert \phi\rangle= \left(2nh+\frac{c}{12}(n^3-n)\right)
=\langle \phi\vert \phi\rangle.
\end{eqnarray}
From setting $n=1$ we see that we must have $h\ge 0$ and from setting $n$ sufficiently large we find that $c \ge 0$. A more detailed and level-by-level analysis~\cite{Friedan:1983xq,Friedan:1986kd,Goddard:1986ee}
shows that a necessary and sufficient condition for an irreducible highest weight representation to be unitary is that either $c \ge 1$ and $h \ge 0$ or that,
\begin{eqnarray}
c = 1 - \frac{6}{m(m+1)} \qquad h=\frac{((m+1)p-mq)^2-1}{4m(m+1)}
\end{eqnarray}
where,
\begin{eqnarray}
m \ge 2\qquad 1 \le p <m-1  \qquad 1 \le p \le q.
\end{eqnarray}
This includes the case $c=h=0$ where the representation is trivial. 
\subsubsection{Further Constraints and Extensions}
There are some further consistency constraints to be taken
into account, such as the elimination of negative-norm states
from the physical spectrum as well as the extension to
supersymmetry. While additional constraints will not be important for our
purposes, it should be noted that they exist.
\subsection{String Theory and the Poincar\'e Algebra}
The string world-sheet is embedded into a higher-dimensional Minkowski space. The Poincar\'e algebra appeared earlier in eq.~(\ref{eq:poincarealg}). In string theory, one has multiple copies of the algebra.
This is not entirely unfamiliar, after all the Fock space for multi-particle theories  is a direct sum of tensor products of single-particle Hilbert spaces. From the string action, standard textbooks~\cite{Witten:1998qj} derive the following
operator representations which satisfy the Poincar\'e algebra: 
\begin{eqnarray}
M^{\mu\nu}=x^{\mu}p^{\nu}-x^{\nu}p^{\mu}-i\sum_{n=1}^{\infty}\frac{1}{n}(\alpha_{-n}^{\mu}\alpha_{n}^{\nu}-\alpha_{-n}^{\nu}\alpha_{n}^{\mu}).
\end{eqnarray}
Here $x^{\mu}p^{\nu}-x^{\nu}p^{\mu}$ together with $p^{\mu}$ satisfy the Poincar\'e algebra. In addition the terms $-i(\alpha_{-n}^{\mu}\alpha_{n}^{\nu}-\alpha_{-n}^{\nu}\alpha_{n}^{\mu})$ for each $n\in \mathbb{N}$ individually are operator representations of the Lorentz algebra. The expression above combines them into a single angular momentum operator where the weights $\frac{1}{n}$ account for the non-standard normalization of the modes in Eq.~(\ref{eq:modes}).
At the same time the modes $\alpha_{k}^{\mu}$ are building blocks for
the Virasoro generators. The Virasoro algebra is independent of the Poincar\'e algebra in the sense that $[L_k,M^{\mu\nu}]=0$.
The symmetry-based approach proposed in this paper has the advantage that the algebra can be more easily modified. For instance it is problematic in string theory that interactions are described by world-sheet topologies over which one has to sum to an infinite series (whose strong-coupling behavior appears to be essential). Instead one can modify the representation. For a start, one expresses $x^{\mu}$ and $p^{\mu}$ in terms of creation and annihilation operators. Further, introduce continuous representations as they are known from point particles. In this way nothings stands in the way of building a multi-string Fock space.
\subsection{Orientation Reversal of the Closed-String}
We can define the operator,
\begin{eqnarray}
\Omega(\theta)=\exp\left\{i\frac{\theta}{2} \left(\sum_{n\ne 0} \alpha^{\mu}_{-n}\tilde\alpha_{\mu n}\right)\right\}.
\end{eqnarray}
It rotates right-movers into left-movers and vice versa,
\begin{eqnarray}
\begin{array}{rcc}
\Omega(\theta) \alpha^{\mu}_k\Omega(\theta)^{-1}&=&\cos(\theta)\alpha_k^{\mu}-\sin(\theta)\tilde\alpha^{\mu}_k\\
\Omega(\theta) \tilde\alpha^{\mu}_k\Omega(\theta)^{-1}&=&-\sin(\theta)\alpha^{\mu}_k+\cos(\theta)\tilde\alpha_k^{\mu}.\\
\end{array}
\end{eqnarray}
The operator $\Omega(-\pi)$ satisfies $\Omega(-\pi)=\Omega^{\dagger}(-\pi)=\Omega^{-1}(-\pi)$ and therefore is an involution $\Omega(-\pi)^2=1$, 
\begin{eqnarray}
\Omega(-\pi) \alpha^{\mu}_k\Omega(-\pi)^{-1}=\tilde{\alpha}_k^{\mu} \qquad
\Omega(-\pi) \tilde{\alpha}^{\mu}_k\Omega(-\pi)^{-1}=\alpha_k^{\mu}.
\end{eqnarray}
It is the familiar orientation reversal operator for the closed string.
Since the generator is not part of the symmetry algebra, requiring closed string orientation invariance amounts to imposing an additional symmetry.
This is a symmetry which ties together the two independent copies of the Virasoro algebra.
\subsection{Orientation Reversal of the Open-String}
In analogy to Eq.~(\ref{eq:numop}) we can define a unitary operator,
\begin{eqnarray}
\Omega(\theta)=e^{i\theta L_0}=\exp\left\{i\theta \left(\sum_{n=1}^{\infty} \alpha^{\mu}_{-n}\alpha_{\mu n}+\frac{1}{2}\alpha^{\mu}_{0}\alpha_{\mu 0}\right)\right\}.
\end{eqnarray}
It acts according to,
\begin{eqnarray}
\Omega(\theta) \alpha^{\mu}_k\Omega(\theta)^{-1}=e^{-ik\theta}\alpha_k^{\mu}.
\end{eqnarray}
For $\theta=\pi$ it reduces to the orientation reversal operator for the open string,
\begin{eqnarray}
\Omega(\pi) \alpha^{\mu}_k\Omega(\pi)^{-1}=(-1)^{k}\alpha_k^{\mu}.
\end{eqnarray} 
In contrast to the the closed string generator the generator $L_0$ is by default part of the symmetry algebra. Note also that $L_0$ (apart from an irrelevant normal ordering constant) is the Hamiltonian, therefore $\Omega(\tau)=e^{iL_0\tau}$ causes time translations. So we are really dealing with invariance under the  discrete time shift $\tau \rightarrow \tau+\pi$.

Since the generator is the open-string Hamiltonian, a natural question to ask is whether an analogous action for the closed string exists. The mass-shell
conditions for the open and closed strings are,
\begin{eqnarray}
\begin{array}{rcl}
\alpha' M_{\text{open}}^2&=&\sum_{n=1}^{\infty}\alpha^{\mu}_{-n}\alpha_{\mu n}\\
\alpha' M_{\text{closed}}^2&=&2\sum_{n=1}^{\infty}(\alpha^{\mu}_{-n}\alpha_{\mu n}+\tilde\alpha^{\mu}_{-n}\tilde\alpha_{\mu n})
\end{array}
\end{eqnarray}
Due to the extra factor of two in the closed string expression an analogous closed string operator acts trivially on the modes.
\subsection{T-Duality}
\subsubsection{Position-Momentum Duality and the String Wave Equation}
The solutions to the string's equations of motion can be looked up in any introductory text on string theory. For the closed string they are $X^{\mu}(\tau,\sigma)=X^{\mu}_{R}(\tau,\sigma)+X^{\mu}_L(\tau,\sigma)$
with,
\begin{eqnarray}
\begin{array}{rcl}
X^{\mu}_R(\tau,\sigma)&=&\frac{1}{2}x^{\mu}+\frac{1}{2}l^2 p^{\mu}(\tau-\sigma)+\frac{i}{2}l\sum_{n\ne0}\frac{1}{n}\alpha_n^{\mu}e^{-2in(\tau-\sigma)}\\
X^{\mu}_L(\tau,\sigma)&=&\frac{1}{2}x^{\mu}+\frac{1}{2}l^2 p^{\mu}(\tau+\sigma)+\frac{i}{2}l\sum_{n\ne0}\frac{1}{n}\tilde{\alpha}_n^{\mu}e^{-2in(\tau+\sigma)},
\end{array}
\end{eqnarray}
where $l=\frac{1}{\sqrt{2\pi\alpha'}}=\frac{1}{\sqrt{\pi T}}$ and $\alpha_0^{\mu}=\tilde{\alpha}_0^{\mu}=\frac{1}{2}lp^{\mu}$. For the open string similar equations exist.
The string momentum is then defined as $P^{\mu}_{\tau}(\tau,\sigma)=\frac{1}{l^{2}}\dot{X}^{\mu}(\tau,\sigma)$.
Note that the coordinate $X^{\mu}(\tau,\sigma)$ is real and $P^{\mu}_{\tau}(\tau,\sigma)$ is imaginary. With this definition, the canonical commutators resurface,
\begin{eqnarray}
\begin{array}{ccl}
\;[X^{\mu}(\tau,\sigma),P^{\nu}_{\tau}(\tau,\sigma)]&=&i\delta(\sigma-\sigma')\eta^{\mu\nu},\\
\;[X^{\mu}(\tau,\sigma),X^{\nu}(\tau,\sigma)]&=&0,\\
\;[P^{\mu}_{\tau}(\tau,\sigma),P^{\nu}_{\tau}(\tau,\sigma)]&=&0.
\end{array}
\end{eqnarray}
We are seeking a transformation similar to  Eq.~(\ref{eq:numop}). However, no unitary transformation on the modes exists which gives rise to  $(X^{\mu},P^{\mu}_{\tau})\rightarrow (P^{\mu}_{\tau},-X^{\mu})$. In the next paragraph a different transformation
is proposed which is a better analogue to Eq.~(\ref{eq:numop}).
\subsubsection{T-Duality Operator}
The analogue to Eq.~(\ref{eq:numop})
for the string is the unitary operator,
\begin{eqnarray}
U(\theta)=\exp\left\{i\theta \left(\sum_{n=1}^{\infty} \frac{1}{n}\alpha^{\mu}_{-n}\alpha_{\mu n}+\frac{1}{2}\alpha^{\mu}_{0}\alpha_{\mu 0}\right)\right\},
\end{eqnarray}
which generates a rotation by the angle $\theta$ in the complex space associated with each mode $n$. The weights $1/n$ which do not appear in Eq.~(\ref{eq:numop})
serve to correct for the different normalization of the mode operators. The action of this operator on the modes is,
\begin{eqnarray}
U(\theta)\alpha^{\mu}_kU^{-1}(\theta)&=&e^{-i\theta}\alpha^{\mu}_k.
\end{eqnarray}
In Eq.~(\ref{eq:numop}) the angle  $\theta=\pi/2$ effectuated the canonical transformation which swaps position and momentum. Here this angle results in,
\begin{eqnarray}
\alpha^{\mu}_{k} \leftrightarrow -\alpha^{\mu}_{k}.
\end{eqnarray}
This is the action of T-duality on the algebra of the bosonic string.
At the same time we argued that it is the analogue to the 
position-momentum duality (which again is essentially the
wave-particle duality). In the following, we further want to support the view that T-duality should be 
understood as a form of wave-particle duality.
\subsubsection{T-Duality as Fourier Transformation}
The swap between position and momentum-space representations
happens by virtue of a Fourier transform:
\begin{eqnarray}
\psi(p)=\langle \psi \vert p\rangle=\langle\psi \vert x\rangle\langle x\vert p\rangle=\int \psi(x)e^{ipx}dx. 
\end{eqnarray}
If T-duality is to amount to a swap between a position and momentum space picture then the Fourier transform must be compatible with the inversion of radii $R \leftrightarrow 1/R$ under $T$-duality. Let us verify that. One can use the process of periodicization to
express a function $f(x)$ on a domain $[- R/2;R/2]$ by another function $F(x)$ on $\mathbb{R}$,
\begin{eqnarray}
f(x)=\sum_{n\in \mathbb{Z}} F(x+Rn),
\end{eqnarray}
so that,
\begin{eqnarray}
\int_{S^1}f(x)dx=\int_{\mathbb{R}} F(x)dx.
\end{eqnarray}
Doing so allows us to write the periodic function
as a convolution,
\begin{eqnarray}
f(x)=F_0(x) \star \sum_{n\in \mathbb{Z}}\delta(x+ R n).
\end{eqnarray}
In this expression one recognizes the Dirac comb function ${\displaystyle \operatorname{III}_R(x)}\equiv\frac{1}{R}\sum_{n\in \mathbb{Z}}\delta(x+ R n)=\frac{1}{R}\sum_{n\in \mathbb{Z}}e^{2\pi i nx/R}$. It has the property that its periodicity inverts
under Fourier transformation,
\begin{eqnarray}
{\displaystyle \operatorname{III}_R(x)} \overset{\mathcal{F}}{\longleftrightarrow} \frac{1}{R}{\displaystyle \operatorname{III}_{\frac{1}{R}}(p)}.
\end{eqnarray}
Therefore the Fourier transform of our function $f(x)$ is,
\begin{eqnarray}
\hat{f}(p)=\frac{1}{R}\sum_{n\in \mathbb{Z}}\hat{f}_0(n/R)\delta(p+  n/R),
\end{eqnarray}
which shows that $R$ indeed is inverted. The argument can be taken further generalized to backgrounds other than cicles. According to the SYZ conjecture, the generalization of T-duality to general Calabi-Yau manifolds is mirror symmetry~\cite{Hori:740255} (which by extension is also related to the Langlands duality).
Any mirror pair including any given configuration of branes should therefore be related by some kind of Fourier transformation. 
Indeed the Fourier-Mukai transform~\cite{mukai1981,orlov2003} is a categorified Fourier transformation and underlies homological mirror symmetry. It transforms between derived categories of coherent sheaves.
This category has a realization in terms of matrix
factorizations~\cite{Eisenbud,Kontsevich:1994dn}.

\subsubsection{Momentum and Winding Modes}
T-duality exchanges the momentum and winding modes,
\begin{eqnarray}
m \longleftrightarrow w\qquad R \longleftrightarrow \frac{1}{R},
\end{eqnarray}
and in the process a $Dp$ brane stretching along the direction of
T-duality becomes a $Dp-1$ brane. Conversely, a $Dp$-brane orthogonal to the direction of T-duality becomes a $Dp+1$ brane. This can be fully understood in terms of Fourier transformations. One has the following Fourier pairing,
\begin{eqnarray}
\langle x \vert x_0\rangle=\delta(x-x_0) \overset{\text{T-duality}}\longleftrightarrow \langle p \lvert x_0 \rangle=e^{-ipx_0}.
\end{eqnarray}
One one side of the duality one has a D0-brane located at position $x_0$.\ From the uncertainty relation it is known that such a localized
quantum object, whose wave-function is described by a delta function, will have a completely unknown momentum distribution and indeed the probability density $|e^{-ipx_0}|dp=dp$ is uniform. We have argued that under T-duality the momentum
wave-function becomes a new position wave-function. Now our position wave-function
is completely uniform, indicating the object is 'smeared' out over the entire space. No point is preferred over another and we conclude that we now have an extended object located everywhere along the T-duality direction. The D0 brane has turned into a D1 brane. Conversely, a D1-brane extended along some dimension $x$ has equal probability to be found anywhere along that dimension. Then its momentum distribution is localized so the T-dual of the D1 brane is a D0 brane.
One can also have $n$ D0 branes at the same point which are transformed to a quantum object with $n$ units of momentum, interpreted as wrapping the compact dimension $n$ times.
Superpositions of a finite number of such objects are known
in the literature as bound states of branes. Conceivable
are also infinite sums. Then one can no longer speak of D0 and D1 branes
but rather has wave packets, such as Gaussian wave-packets in a non-compact space. As in non-compact space, wave-functions exist which
are self-dual under Fourier transforms~\cite{Fedotowski:1972}.
In that case one can no longer speak of a fixed brane dimension.

The claims concerning T-duality are now briefly summarized. As discussed, T-duality as a geometric generalization of the Fourier
transform is nothing new. But  it is argued here that T-duality should be interpreted as a map between a position space and
momentum space representation. This leads to a rather counter-intuitive
seeming unification of space and momentum.
Position and momentum are the real and the imaginary axes of a complex space -- the phase space. T-duality, like the position-momentum duality, is a rotation by 90 degrees in this phase space. This duality is not a specific features of particular theories,
but rather is inherently grounded in quantum theory. Quantum theory, through its inherent usage of projective representations, defines states
only up to a complex phase. Thereby it smears out the distinction between the real and the imaginary axes of this space. At the same time, this is cause of the uncertainty relation.
\subsection{Interactions with Massless Closed String States}
After the digression to T-duality, we are now reverting to the symmetry based approach of string theory and address introducing interactions. The massless closed states of bosonic string theory are,
\begin{eqnarray}
|\Omega^{ij}\rangle=\alpha_{-1}^{i}\tilde{\alpha}_{-1}^j |0\rangle.
\end{eqnarray}
The traceless symmetric part is the graviton, the anti-symmetric part the analog of the magnetic states and the trace is the dilaton. These states all have their equivalent in the superstring. 
In order to obtain interactions with massless closed string states we define the operator,
\begin{eqnarray}
U(M)=\exp\{M_{\mu\nu} (\alpha_{-1}^{\mu} \tilde{\alpha}_{-1}^{\nu}-\alpha_{1}^{\mu} \tilde{\alpha}^{\nu}_{1})\}.
\end{eqnarray}
We can decompose $M_{\mu\nu}$ into a traceless symmetric part $G_{\mu\nu}$,
an anti-symmetric part $B_{\mu\nu}$ and a trace component $\Phi$. 
If we impose invariance under orientation reversal,
\begin{eqnarray}
[\Omega, U(M_{\mu\nu})]=0,
\end{eqnarray}
then $M_{\mu\nu}$ must be symmetric and it follows that the unoriented string theories do not support a $B_{\mu\nu}$ field. Using the algebra and the Baker-Hausdorff Lemma we derive that the operator generates the following deformation:
\begin{eqnarray}
\begin{array}{rcl}
U(M)a_{-1}^{\kappa}U(M)^{-1}&=&(\eta^{\kappa}{}_{\mu_1}+\frac{1}{2!}M{^{\kappa}{}_{\mu_2} M_{\mu_1}{}^{\mu_2}}+\frac{1}{4!}M{^{\kappa}{}_{\mu_4} M_{\mu_3}{}^{\mu_4}}M^{\mu_3}{}_{\mu_2}M_{\mu_1}{}^{\mu_2}+..)a_{-1}^{{\mu_1}}\\
&&-(\frac{1}{1!}M{^{\kappa}{}_{\mu_1} +\frac{1}{3!}M{^{\kappa}}{}_{\mu_3} M_{\mu_2}{}^{\mu_3}}M^{\mu_2}{}_{\mu_1}+..)\tilde{a}^{\mu_1}_{1}.
\end{array}
\end{eqnarray}
For symmetric $M_{\mu\nu} \equiv G_{\mu\nu} $ one has $M_{\mu_i}{}^{\mu_j}=M^{\mu_j}{}_{\mu_i}$ and similarly for the trace component, whereas for anti-symmetric $M_{\mu\nu}\equiv B_{\mu\nu} $  one has $M_{\mu_i}{}^{\mu_j}=-M^{\mu_j}{}_{\mu_i}$. The deformation becomes,
\begin{eqnarray}
\begin{array}{rcl}
U(G)\alpha_{-1}^{\kappa}U(G)^{-1}&=&\cosh(G^{\kappa}{}_{\mu})\alpha^{\mu}_{-1}-\sinh(G^{\kappa}{}_{\mu})\tilde{\alpha}^{\mu}_{1}\\
U(\Phi)\alpha_{-1}^{\kappa}U(\Phi)^{-1}&=&\cosh(\Phi)\alpha^{\mu}_{-1}-\sinh(\Phi)\tilde{\alpha}^{\mu}_{1}\\
U(B)\alpha_{-1}^{\kappa}U(B)^{-1}&=&\cos(B^{\kappa}{}_{\mu})\alpha^{\mu}_{-1}-\sin(B^{\kappa}{}_{\mu})\tilde{\alpha}^{\mu}_{1}.
\end{array}
\end{eqnarray}
The corresponding expressions for the other mode operators are listed in the appendix. The appendix also contains more detailed intermediate steps of some of the subsequent calculations.
\subsection{Symmetric Deformation}
The deformation of the only non-trivial term in the Hamiltonian $N+\bar{N}-2$ is given by,
\begin{eqnarray}
\begin{array}{rcl}
&&U(G)(\alpha^{\mu}_{-1}\alpha_{1\mu}+\tilde{\alpha}^{\mu}_{-1}\tilde{\alpha}_{1\mu})U(G)^{-1}
\\
&=&\alpha^{\mu}_{-1}\alpha_{1\mu}+\tilde{\alpha}^{\mu}_{-1}\tilde{\alpha}_{1\mu}
+(\tilde{\alpha}^{\mu}_{-1}\alpha^{\nu}_{-1}
+\tilde{\alpha}^{\mu}_{1}\alpha^{\nu}_{1})\sinh (2G_{\mu\nu})
+(\eta_{\mu\nu}-\cosh(2G_{\mu\nu}))\eta^{\mu\nu}.
\end{array}
\end{eqnarray}
The relevant component of the level-matching operator $N-\bar{N}$ transforms according to,
\begin{eqnarray}
U(G)(\alpha^{\mu}_{-1}\alpha_{1\mu}-\tilde{\alpha}^{\mu}_{-1}\tilde{\alpha}_{1\mu})U(G)^{-1}
=(\alpha^{\mu}_{-1}\alpha^{\nu}_{1}-\tilde{\alpha}^{\mu}_{-1}\tilde{\alpha}^{\nu}_{1})\cosh (2G_{\mu\nu}).
\end{eqnarray}
Therefore, at first order, the Hamiltonian acquires the interaction term,
\begin{eqnarray}
\Delta H=2G_{\mu\nu}(\tilde{\alpha}^{\mu}_{-1}\alpha^{\nu}_{-1}
+\tilde{\alpha}^{\mu}_{1}\alpha^{\nu}_{1}).
\end{eqnarray}
The level matching condition remains unchanged at lowest order.
Let us return to the full deformation and substitute $G=\ln(E)$. Since $G$ is an invertible symmetric matrix the matrix $E$ is also symmetric.  Using, 
\begin{eqnarray}
\begin{array}{cc}
2\cosh(G)=E+E^{-1} & \qquad2\sinh(G)=E-E^{-1}\\
2\cosh(2G)=E^2+E^{-2} & \qquad2\sinh(2G)=E^{2}-E^{-2},
\end{array}
\end{eqnarray}
one gets,
\begin{eqnarray}
\begin{array}{rcl}
&&U(G)(\alpha^{\mu}_{-1}\alpha_{1\mu}+\tilde{\alpha}^{\mu}_{-1}\tilde{\alpha}_{1\mu})U(G)^{-1}
\\
&=&\alpha^{\mu}_{-1}\alpha_{1\mu}+\tilde{\alpha}^{\mu}_{-1}\tilde{\alpha}_{1\mu}
+\frac{1}{2}(\tilde{\alpha}^{\mu}_{-1}\alpha^{\nu}_{-1}
+\tilde{\alpha}^{\mu}_{1}\alpha^{\nu}_{1})(E^{2}-E^{-2})_{\mu\nu}
-\frac{1}{2}\text{Tr} (E^{2}+E^{-2})+(d-2).
\end{array}
\end{eqnarray}
We derived a Hamiltonian which at all orders exhibits the
duality,
\begin{eqnarray}
E \leftrightarrow E^{-1} \qquad \alpha^{\mu}_{k} \leftrightarrow -\alpha^{\mu}_{k}.
\end{eqnarray}
For the dilaton the calculation is almost identical with the matrix $G$ replaced with the scalar $\Phi$. The duality is,
\begin{eqnarray}
e^{\Phi} \leftrightarrow e^{-{\Phi}} \qquad \alpha^{\mu}_{k} \leftrightarrow -\alpha^{\mu}_{k}.
\end{eqnarray}
Using $g_s\equiv e^{\langle \Phi \rangle}$ the first mapping is that of S-duality $g_s \leftrightarrow g_s^{-1}$ whereas the second is T-duality, that is, the Hamiltonian possesses U-duality. 
\subsection{Anti-Symmetric Deformation}
The deformation of the only non-trivial term in the Hamiltonian $N+\bar{N}+2$  under the anti-symmetric action is given by,
\begin{eqnarray}
\begin{array}{rcl}
&&U(B)(\alpha^{\mu}_{-1}\alpha_{1\mu}+\tilde{\alpha}^{\mu}_{-1}\tilde{\alpha}_{1\mu})U(B)^{-1}
\\
&=&(\alpha^{\mu}_{-1}\alpha^{\nu}_{1}+\tilde{\alpha}^{\mu}_{-1}\tilde{\alpha}^{\nu}_{1})\cos(2B_{\mu\nu})
+(\tilde{\alpha}^{\mu}_{-1}\alpha^{\nu}_{-1}
-\tilde{\alpha}^{\mu}_{1}\alpha^{\nu}_{1})\sin(2B_{\mu\nu})
+(\eta_{\mu\nu}-\cos(2B_{\mu\nu}))\eta^{\mu\nu}.
\end{array}
\end{eqnarray}
We arrived at a Hamiltonian which is periodic in the $B_{\mu\nu}$ field. The level-matching operator $N-\bar{N}$ remains invariant at all orders,
\begin{eqnarray}
U(B)(\alpha^{\mu}_{-1}\alpha_{1\mu}-\tilde{\alpha}^{\mu}_{-1}\tilde{\alpha}_{1\mu})U(B)^{-1}
=\alpha^{\mu}_{-1}\alpha_{1\mu}-\tilde{\alpha}^{\mu}_{-1}\tilde{\alpha}_{1\mu}.
\end{eqnarray}
This means that the influence of the $B_{\mu\nu}$ field cannot break closed strings apart. At lowest order the Hamiltonian acquires an interaction term,
\begin{eqnarray}
\Delta H=2B_{\mu\nu}(\tilde{\alpha}^{\mu}_{-1}\alpha^{\nu}_{-1}
-\tilde{\alpha}^{\mu}_{1}\alpha^{\nu}_{1}).
\end{eqnarray}
There are many possible avenues for actual computations based on this work
but they are planned be part of subsequent papers. The purpose of this work was only to lay the necessary foundations. 

%% file: StringBgr_appendix.tex
\appendix
\section{Action of the Deformation Operators}
\subsection{Anti-symmetric deformation}
The anti-symmetric deformation operator,
\begin{eqnarray}
U(B)=\exp\{B_{\mu\nu} (\alpha_{-1}^{\mu} \tilde{\alpha}_{-1}^{\nu}-\alpha_{1}^{\mu} \tilde{\alpha}^{\nu}_{1})\},
\end{eqnarray}
acts on the modes according to,
\begin{eqnarray}
\begin{array}{ccr}
U(B)\alpha_{-1}^{\kappa}U(B)^{-1}&=&\cos(B^{\kappa}{}_{\mu})\alpha^{\mu}_{-1}-\sin(B^{\kappa}{}_{\mu})\tilde{\alpha}^{\mu}_{1}\\
U(B)\alpha_{ 1}^{\kappa}U(B)^{-1}&=&-\sin(B^{\kappa}{}_{\mu})\tilde{\alpha}^{\mu}_{-1}+\cos(B^{\kappa}{}_{\mu})\alpha^{\mu}_{1},
\\\\
U(B)\tilde{\alpha}_{-1}^{\kappa}U(B)^{-1}&=&\cos(B^{\kappa}{}_{\mu})\tilde{\alpha}^{\mu}_{-1}+\sin(B^{\kappa}{}_{\mu})\alpha^{\mu}_{1}\\
U(B)\tilde{\alpha}_{ 1}^{\kappa}U(B)^{-1}&=&\sin(B^{\kappa}{}_{\mu})\alpha^{\mu}_{-1}+\cos(B^{\kappa}{}_{\mu})\tilde{\alpha}^{\mu}_{1},
\end{array}
\end{eqnarray} 
It follows: 
\begin{eqnarray}
\begin{array}{rcl}
&&U(B)\alpha^{\kappa}_{-1}\alpha_{\kappa 1}U(B)^{-1}
\\
&=&(\cos(B^{\kappa}{}_{\mu})\alpha^{\mu}_{-1}-\sin(B^{\kappa}{}_{\mu})\tilde{\alpha}^{\mu}_{1})
(-\sin(B_{\kappa}{}_{\nu})\tilde{\alpha}^{\nu}_{-1}+\cos(B_{\kappa}{}_{\nu})\alpha^{\nu}_{1})
\\\\
&=&\cos(B^{\kappa}{}_{\mu})\cos(B_{\kappa}{}_{\nu})\alpha^{\mu}_{-1}\alpha^{\nu}_{1}+
\sin(B^{\kappa}{}_{\mu})\sin(B_{\kappa}{}_{\nu})\tilde{\alpha}^{\mu}_{1}\tilde{\alpha}^{\nu}_{-1}
\\
&&-\cos(B^{\kappa}{}_{\mu})\sin(B_{\kappa}{}_{\nu})\tilde{\alpha}^{\mu}_{-1}\alpha^{\nu}_{-1}
-\sin(B^{\kappa}{}_{\mu})\cos(B_{\kappa}{}_{\nu})\tilde{\alpha}^{\mu}_{1}\alpha^{\nu}_{1}
\\\\
&=&\cos(B_{\mu}{}^{\kappa})\cos(B_{\kappa}{}_{\nu})\alpha^{\mu}_{-1}\alpha^{\nu}_{1}-
\sin(B_{\mu}{}^{\kappa})\sin(B_{\kappa}{}_{\nu})(\tilde{\alpha}^{\nu}_{-1}\tilde{\alpha}^{\mu}_{1}+\eta^{\mu\nu})
\\
&&-\cos(B_{\mu}{}^{\kappa})\sin(B_{\kappa}{}_{\nu})\tilde{\alpha}^{\mu}_{-1}\alpha^{\nu}_{-1}
+\sin(B_{\mu}{}^{\kappa})\cos(B_{\kappa}{}_{\nu})\tilde{\alpha}^{\mu}_{1}\alpha^{\nu}_{1}
\\\\
&=&\frac{1}{2}(\eta_{\mu\nu}+\cos(2B_{\mu\nu}))\alpha^{\mu}_{-1}\alpha^{\nu}_{1}-
\frac{1}{2}(\eta_{\mu\nu}-\cos(2B_{\mu\nu}))(\tilde{\alpha}^{\nu}_{-1}\tilde{\alpha}^{\mu}_{1}+\eta^{\mu\nu})
-\frac{1}{2}\sin(2B_{\mu\nu})(\tilde{\alpha}^{\mu}_{-1}\alpha^{\nu}_{-1}
-\tilde{\alpha}^{\mu}_{1}\alpha^{\nu}_{1})
\end{array}\nonumber
\end{eqnarray}
Similarly we find,
\begin{eqnarray}
\begin{array}{rcl}
&&U(B)\tilde{\alpha}^{\kappa}_{-1}\tilde{\alpha}_{\kappa 1}U(B)^{-1}
\\
&=&(\cos(B^{\kappa}{}_{\mu})\tilde{\alpha}^{\mu}_{-1}+\sin(B^{\kappa}{}_{\mu})\alpha^{\mu}_{1})
(\sin(B_{\kappa}{}_{\nu})\alpha^{\nu}_{-1}+\cos(B_{\kappa}{}_{\nu})\tilde{\alpha}^{\nu}_{1})
\\
&=&\frac{1}{2}(\eta_{\mu\nu}+\cos(2B_{\mu\nu}))\tilde{\alpha}^{\mu}_{-1}\tilde{\alpha}^{\nu}_{1}-
\frac{1}{2}(\eta_{\mu\nu}-\cos(2B_{\mu\nu}))(\alpha^{\mu}_{-1}\alpha^{\nu}_{1}+\eta^{\mu\nu})
-\frac{1}{2}\sin(2B_{\mu\nu})(\tilde{\alpha}^{\mu}_{-1}\alpha^{\nu}_{-1}
-\tilde{\alpha}^{\mu}_{1}\alpha^{\nu}_{1}).
\end{array}\nonumber
\end{eqnarray}
\subsection{Symmetric deformation}
The deformation with the symmetric deformation operator differs from its anti-symmetric counterpart in that the trigonometric functions are replaced by their hyperbolic equivalents. The symmetric deformation operator,
\begin{eqnarray}
U(G)=\exp\{G_{\mu\nu} (\alpha_{-1}^{\mu} \tilde{\alpha}_{-1}^{\nu}-\alpha_{1}^{\mu} \tilde{\alpha}^{\nu}_{1})\},
\end{eqnarray}
acts on the mode operators according to,
\begin{eqnarray}
\begin{array}{ccr}
U(G)\alpha_{-1}^{\kappa}U(G)^{-1}&=&\cosh(G^{\kappa}{}_{\mu})\alpha^{\mu}_{-1}-\sinh(G^{\kappa}{}_{\mu})\tilde{\alpha}^{\mu}_{1}\\
U(G)\alpha_{ 1}^{\kappa}U(G)^{-1}&=&-\sinh(G^{\kappa}{}_{\mu})\tilde{\alpha}^{\mu}_{-1}+\cosh(G^{\kappa}{}_{\mu})\alpha^{\mu}_{1}.
\end{array}
\end{eqnarray} 
The action on the anti-holomorphic operators is identical with the holomorphic and anti-holomorphic modes swapped.
From these expressions we find,
\begin{eqnarray}
\begin{array}{rcl}
&&U(G)\alpha^{\kappa}_{-1}\alpha_{\kappa 1}U(G)^{-1}
\\
&=&(\cosh (G^{\kappa}{}_{\mu})\alpha^{\mu}_{-1}-\sinh(G^{\kappa}{}_{\mu})\tilde{\alpha}^{\mu}_{1})
(-\sinh(G_{\kappa}{}_{\nu})\tilde{\alpha}^{\nu}_{-1}+\cosh(G_{\kappa}{}_{\nu})\alpha^{\nu}_{1})
\\\\
&=&\cosh(G^{\kappa}{}_{\mu})\cosh(G_{\kappa}{}_{\nu})\alpha^{\mu}_{-1}\alpha^{\nu}_{1}+
\sinh(G^{\kappa}{}_{\mu})\sinh(G_{\kappa}{}_{\nu})\tilde{\alpha}^{\mu}_{1}\tilde{\alpha}^{\nu}_{-1}
\\
&&-\cosh(G^{\kappa}{}_{\mu})\sinh(G_{\kappa}{}_{\nu})\tilde{\alpha}^{\mu}_{-1}\alpha^{\nu}_{-1}
-\sinh(G^{\kappa}{}_{\mu})\cosh(G_{\kappa}{}_{\nu})\tilde{\alpha}^{\mu}_{1}\alpha^{\nu}_{1}
\\\\
&=&\cosh (G_{\mu}{}^{\kappa})\cosh(G_{\kappa}{}_{\nu})\alpha^{\mu}_{-1}\alpha^{\nu}_{1}+
\sinh(G_{\mu}{}^{\kappa})\sinh(G_{\kappa}{}_{\nu})(\tilde{\alpha}^{\nu}_{-1}\tilde{\alpha}^{\mu}_{1}+\eta^{\mu\nu})
\\
&&-\cosh(G_{\mu}{}^{\kappa})\sinh(G_{\kappa}{}_{\nu})\tilde{\alpha}^{\mu}_{-1}\alpha^{\nu}_{-1}
-\sinh(G_{\mu}{}^{\kappa})\cosh(G_{\kappa}{}_{\nu})\tilde{\alpha}^{\mu}_{1}\alpha^{\nu}_{1}
\\\\
&=&\frac{1}{2}(\eta_{\mu\nu}+\cosh (2G_{\mu\nu}))\alpha^{\mu}_{-1}\alpha^{\nu}_{1}+
\frac{1}{2}(\eta_{\mu\nu}-\cosh (2G_{\mu\nu}))(\tilde{\alpha}^{\nu}_{-1}\tilde{\alpha}^{\mu}_{1}+\eta^{\mu\nu})\\
&&-\frac{1}{2}\sinh (2G_{\mu\nu})(\tilde{\alpha}^{\mu}_{-1}\alpha^{\nu}_{-1}
+\tilde{\alpha}^{\mu}_{1}\alpha^{\nu}_{1}).
\end{array}
\end{eqnarray}
Similarly we find,
\begin{eqnarray}
\begin{array}{rcl}
&&U(G)\tilde{\alpha}^{\mu}_{-1}\tilde{\alpha}_{\mu 1}U(G)^{-1}
\\
&=&(\cosh(G^{\kappa}{}_{\mu})\tilde{\alpha}^{\mu}_{-1}-\sinh(G^{\kappa}{}_{\mu})\alpha^{\mu}_{1})
(-\sinh(G_{\kappa}{}_{\nu})\alpha^{\nu}_{-1}+\cosh(G_{\kappa}{}_{\nu})\tilde{\alpha}^{\nu}_{1})
\\
&=&\frac{1}{2}(\eta_{\mu\nu}+\cosh (2G_{\mu\nu}))\tilde{\alpha}^{\mu}_{-1}\tilde{\alpha}^{\nu}_{1}+
\frac{1}{2}(\eta_{\mu\nu}-\cosh(2G_{\mu\nu}))(\alpha^{\mu}_{-1}\alpha^{\nu}_{1}+\eta^{\mu\nu})\\
&&-\frac{1}{2}\sinh (2G_{\mu\nu})(\tilde{\alpha}^{\mu}_{-1}\alpha^{\nu}_{-1}
+\tilde{\alpha}^{\mu}_{1}\alpha^{\nu}_{1}).
\end{array}
\end{eqnarray}